\documentclass[numbook,natbib,final,runningheads]{svjour3}
\usepackage{graphicx,mathptmx} 
\usepackage{amssymb} 
\usepackage{makeidx} 
\makeindex

\newcommand{\hsi}{\textit{RHESSI}}

\newcommand{\NRH}{Nan{\c c}ay Radio Heliograph}

\newcommand{\kmps}{km s${^{-1}}$}

\newcommand{\pccm}{cm${^{-3}}$}
\newcommand{\degr}{\ifmmode^\circ\else$^\circ$\fi}
\newcommand{\arcsec}{\hbox{$^{\prime\prime}$}}

\begin{document}

\title{The Relationship Between Solar Radio and Hard X-ray Emission}

\author{S.~M.~White$^{1,5}$,
A.~O. Benz$^{2}$,
S.~Christe$^{3}$,
F.~F{\'a}rn{\'i}k$^{4}$,
M.~R.~Kundu$^{5}$,
G.~Mann$^{6}$,
Z.~Ning$^{7}$,
J.-P.~Raulin$^{8}$,
A.~V.~R.~Silva-V{\'a}lio$^{8}$,
P.~Saint-Hilaire$^{9}$,
N.~Vilmer$^{10}$, and
A.~Warmuth$^{6}$
}

\institute{
$^{1}$Space Vehicles Directorate, AFRL, Kirtland AFB, Albuquerque, NM 87123 USA; \email{Stephen.White@kirtland.af.mil}\\
$^2$Institute of Astronomy, ETH Z{\"u}rich, 8093 Z{\"u}rich, Switzerland \\
$^3$NASA/Goddard Space Flight Center, Mail Code 671, Greenbelt, MD 20771, USA \\
$^4$Astronomical Institute, Academy of Sciences, 251 65 Ond\v{r}ejov, Czech Republic \\
$^5$Dept. of Astronomy, University of Maryland, College Park, MD 20742 USA \\
$^{6}$Astrophysikalisches Institut Potsdam, An der Sternwarte 16, D 14482 Potsdam, Germany \\
$^7$Purple Mountain Observatory, Nanjing 210008, China \\
$^8$Centro de Radio Astronomia e Astrofisica Mackenzie, R. da
Consola\c{c}\~{a}o 896, 01302-907, S\~{a}o Paulo, SP, Brazil \\
$^{9}$Space Sciences Laboratory, University of California, Berkeley, CA 94720, USA \\
$^{10}$Observatoire de Paris, LESIA – CNRS UMR 8109, 5 Place Jules Janssen, 92190 Meudon, France \\
             }
             
\authorrunning{S. M. White et al.}
\titlerunning{Solar radio and hard X-ray emission}
\maketitle

\begin{abstract}

This review discusses the complementary relationship between 
radio and hard X-ray observations of the Sun
using primarily results from the era of the \textit{Reuven Ramaty High Energy
Solar Spectroscopic Imager} satellite.
A primary focus of joint radio and
hard X-ray studies of solar flares uses observations of nonthermal
gyrosynchrotron emission at radio wavelengths and bremsstrahlung hard
X-rays to study the properties of electrons accelerated in the main
flare site, since it is well established that these two emissions show
very similar temporal behavior. A quantitative prescription is
given for comparing the electron energy distributions derived separately
from the two wavelength ranges: this is an important application with
the potential for measuring the magnetic field strength in the flaring
region, and reveals significant differences between the electrons in
different energy ranges. 
Examples of the use of simultaneous data from the two wavelength ranges 
to derive physical conditions are then discussed, including the case of
microflares,\index{microflares} and the comparison of
images at radio and hard X-ray wavelengths is presented. There have
been puzzling results obtained from observations of solar flares at
millimeter and submillimeter wavelengths, and the comparison of these
results with corresponding hard X-ray data is presented.
Finally, the review discusses the association of 
hard X-ray releases with radio emission at
decimeter and meter wavelengths, which is dominated by plasma emission
(at lower frequencies) and electron cyclotron maser emission (at higher
frequencies), both coherent emission mechanisms\index{radio emission!coherent!high efficiency} that require small
numbers of energetic electrons. These comparisons show broad general
associations but detailed correspondence remains more elusive.

\end{abstract}

\setcounter{tocdepth}{3}
\tableofcontents

\section{Introduction}\label{sec:white-1}

From its earliest days, the study of hard X-ray emission from the Sun has had 
a natural ally in solar radio emission \citep{Kun61a}. 
The reason is straightforward: the electrons that produce
hard X-ray emission, by definition, have energies of order 10~keV or more,
and such energetic electrons are also very efficient emitters of radio
emission in the solar corona. That they can produce radio emission 
by a number of different physical mechanisms, in contrast to the
bremsstrahlung-dominated hard X-rays, means that the radio data provide a
range of diagnostics that complement the hard X-ray measurements.
Between these two wavelength ranges we should expect to be sensitive to most
sources of energetic electrons, both thermal and nonthermal, in the solar
corona.

In this article we will review developments in our understanding of the
joint use of radio and hard X-ray emission since the launch of the \textit{Reuven Ramaty High
Energy Solar Spectroscopic Imager} (\hsi) satellite in 2002.
In all cases discussed here, the hard X-ray emission most likely arises from
bremsstrahlung emitted when energetic electrons are accelerated by Coulomb
forces in collisions with ambient ions,
either in the chromosphere or in the corona. 
Bremsstrahlung hard X-ray emission is proportional to the
product of the nonthermal electron density and the ambient ion density.
Inverse Compton emission is
also a possible emission mechanism for the hard X-rays, but more extreme
conditions are required and are not believed to be relevant for our
discussion \citep[e.g., see ][]{KBC08}.\index{inverse Compton radiation}
There are two general areas in which comparison of radio and hard X-ray (HXR)
emission is relevant: phenomena (usually at higher radio frequencies,
i.e., above 5~GHz) 
\index{radio emission!incoherent!and HXRs}
\index{radio emission!overlap with X-ray diagnostics}
\index{hard X-rays!overlap with radio diagnostics}
in which the radio emission is incoherent
gyrosynchrotron emission\footnote{A note on terminology: it is
conventional to reserve the term ``synchrotron'' for emission by highly
relativistic electrons (Lorentz factor $\gamma\,\gg\,1$) spiraling
along magnetic field lines, while emission by mildly relativistic
electrons is referred to as ``gyrosynchrotron'' in solar radiophysics.}, 
and those (usually at lower frequencies, i.e., below 2 GHz) 
where the radio emission is due to a
\index{radio emission!incoherent!gyrosynchrotron emission}\index{radio emission!coherent!plasma emission}\index{radio emission!incoherent}\index{radio emission!coherent!electron cyclotron maser}\index{synchrotron emission!contrasted with gyrosynchrotron emission}
coherent emission mechanism such as plasma emission or cyclotron maser\index{gyrosynchrotron emission!definition}.

\section{Quantitative comparison of hard X-ray and microwave spectra}\label{sec:white-2}

The study of radio emission from solar flares predates the detection of
hard X-rays from the Sun, and the presence of activity at microwave
frequencies in the impulsive phase of solar flares had been well established
by 1958 when the first observation of a flare at hard
X-ray/$\gamma$-ray wavelengths was carried out with a balloon-borne
telescope over Cuba \citep{PeW58}. 
The authors of that paper noted the 
coincidence of the HXRs with microwave emission, and speculated that the
``radio spectrum is emitted as betatron radiation\index{betatron radiation} from [electrons at 
energies of about 1~MeV] as they spiral in the intense magnetic fields
associated with the sunspot group\index{sunspots!and history of X-ray astronomy}.
These same electrons finally stop in
the solar photosphere, where a small fraction of their energy is lost as
bremsstrahlung $\gamma$-rays'' \citep[see also][]{Kun61a}.
This remains our overall picture of the
relationship between hard X-ray and microwave emission in the impulsive
phase of flares: electrons with energies above 30 keV produce gyrosynchrotron 
\index{radio emission!gyrosynchrotron!electron energies}
radio emission as they spiral in ubiquitous 
\index{magnetic field!ubiquity in corona}
magnetic fields in the corona, and bremsstrahlung HXRs
when they precipitate into the dense chromosphere and lose their energy
through collisions. The magnetic field
strength $B$ in the corona is typically of order 500~G over
active regions, leading to electron gyrofrequencies ($2.8\,\times\,10^6\,B$
Hz) of order 1-2~GHz in the low
corona, and electrons with energies of order 100~keV and above radiate
gyrosynchrotron emission at harmonics from the third upwards, usually dominating radio
\index{radio emission!gyrosynchrotron!harmonics}
emission from flares at frequencies above 3~GHz.\index{active regions!high Alfv{\' e}n speed}
Typical flare radio
spectra peak at 10~GHz \citep{GuC75}, although in larger flares the spectral 
peak can often occur at much higher frequencies.

In this picture we expect images of the radio emission at microwave
frequencies to outline the field lines in the corona 
occupied by nonthermal electrons, while the HXRs show the same electron
population as they strike the chromosphere at the footpoints of those field
lines. With sufficiently good imaging, one could establish field-line
connectivity by such a comparison. This picture can be distorted in
several ways: the strength of the radio emission is proportional to a
high power of $B$ \citep[e.g.,][]{DuM82} and thus electrons trapped in
regions of strong magnetic fields will appear very bright, whereas
strong magnetic fields near the surface tend to prevent precipitation of
electrons with non-zero pitch angles due to magnetic mirroring and this
effect could actually reduce gyrosynchrotron emission close to the
footpoints compared to the rest of the loop\index{spectrum!gyrosynchrotron!footpoints}\index{footpoints!gyrosynchrotron emission}.

Such comparisons of gyrosynchrotron and HXR emission 
remain a staple of high energy solar physics.
One aspect to be borne in mind is that although radio telescopes are
capable of excellent imaging and are generally more sensitive to
nonthermal electrons than are HXR detectors, solar-dedicated radio
telescopes capable of excellent imaging are few in number and 
it is difficult to arrange observations during flaring periods on 
non-solar-dedicated
major facilities such as the Very Large Array (VLA). 
Therefore most of the
radio work at microwave frequencies utilizes data from solar-dedicated
arrays such as the Nobeyama
Radio Heliograph (NoRH), the Owens Valley Solar Array (OVSA) or the
Siberian Solar Radio Telescope.
\index{observatories!Very Large Array (VLA)}
\index{observatories!Owens Valley Solar Array (OVSA)}
\index{observatories!Nobeyama Radio Heliograph (NoRH)}
\index{observatories!Siberian Solar Radio Telescope (SSRT)}

We discuss the formalism for quantitative comparison of gyrosynchrotron radio 
data and bremsstrahlung hard X-ray measurements in the current section. 
Examples of scientific topics addressed specifically by comparison of
\textit{\textit{RHESSI}} and microwave observations are given in the following sections.
\index{satellites!RHESSI@\textit{RHESSI}}\index{RHESSI@\textit{RHESSI}}

\subsection{Hard X-ray bremsstrahlung spectra}

In the following discussion of HXR emission we follow the prescription of \citet{HCK78},
based in turn on \citet{Bro71} and the Bethe-Heitler form for the
bremsstrahlung cross-section \citep[see the discussion of
cross-sections in][]{KBE10}. 
\index{bremsstrahlung!Bethe-Heitler cross-section}
Essentially the same calculation was
used by \citet{NWS91} and \citet{WKS03} for quantitative comparison of
radio- and HXR-emitting electrons.
Assume that the hard X-ray photon spectrum 
is a power law of the form 

\begin{equation}
\Phi(\epsilon)\,=\,A_0\,\biggl({\epsilon \over E_0}\biggr)^{-\gamma}
\ \ \ \ {\rm photons\ cm^{-2}\ s^{-1}\ keV^{-1}}
\label{eqn:white-1}
\end{equation}

\noindent where $\epsilon$ is the photon energy,
$\gamma$ is the power-law slope of the photon spectrum, and $A_0$
is the normalization constant at a fiducial photon energy $E_0$ keV. In
the case of the widely-used 
OSPEX package of Richard Schwartz, $E_0$ = 50 keV is the default for the
broken power-law fit to the spectrum and the normalization obtained
from the fit refers to this energy. Given this photon spectrum, the
corresponding electron flux energy spectrum into the target is given by

\begin{equation}
{{d^2 {\cal N}(E)} \over {dE\,dt}}\ =\
3.28\,\times\,10^{33}\ {{A_0\ b(\gamma)} \over E_{0,{\rm keV}}} \ \biggl({E \over
E_0}\biggr)^{-(\gamma+1)} \ \ \ \ {\rm electrons\ keV^{-1}\ s^{-1}}
\label{eqn:white-2}
\end{equation}

\noindent \citep{HCK78}, where $b({\gamma})\ =\ {\gamma}^2\ ({\gamma}-1)^2\ B({\gamma}-0.5,1.5)$ and
$B(x,y)$ is the beta function.
The parameter $b({\gamma})$ is of order 10 to 60 for typical values of
$\gamma$ (3 to 6).  The factor of $E_0$ appearing in the denominator here is
in units of keV \citep[per][p. 498, equation 15]{Bro71}.
We use ${\cal N}(E)$ to denote the total number of electrons at
energy $E$ in some volume: derivatives of this quantity
with respect to volume and energy yield the volume number density and 
the energy distribution, respectively.

\subsection{Thick-target hard X-rays}
\index{thick-target model}\index{hard X-rays!thick-target}

The electron flux into the HXR target cannot be compared directly with the 
microwave parameters since the relevant quantity for the radio emission
is the total number of radiating electrons in the coronal volume, rather than a flux.  
Comparison
of the two depends on the nature of the hard X-ray source. In this section
we consider the so-called ``thick-target'' case, in which the radiating
electrons immediately lose all their energy in a high-density source.  The
density of the source then does not appear in the formulae. To
derive the nonthermal electron volume density from the energy distribution
of the electron flux into the thick target, we assume the classic
expression relating a density to a flux, 
\begin{equation}
{{d^2 {\cal N}(E)} \over {dE\,dt}}\ =\ A_X\, v\, {{d^2 {\cal N}(E)} \over {dE\,dV}}
\label{eqn:white-3}
\end{equation}
where $A_X$ is the area of the X-ray source and $v$ is the
velocity of the electrons into the target.  
$A_X$ must be obtained from observations,
and its determination relies on high-quality imaging data. 

The velocity $v$ is treated as follows. For ultrarelativistic
electrons it is assumed to be of order $c$ with no energy dependence.
For nonrelativistic electrons we assume, as an approximation, that the
component of velocity into the target (i.e., the downwards component for
thick-target emission from the chromosphere) 
carries one-third of the electron energy
(equipartition between directions of motion). Thus we set
\begin{equation}
{1 \over 2}mv^2\ = {1 \over 3}\,E
\label{eqn:white-4}
\end{equation}
\noindent which leads to
\begin{equation}
v\ =\ c\ \sqrt{{2 \over 3}\,{E \over mc^2}}\ =\ 0.0361\ E_{\rm
keV}^{0.5}\ c\ =\ 1.08\,\times\,10^9\ E_{\rm keV}^{0.5} \ \ \ \ {\rm cm\ s^{-1}}.
\label{eqn:white-5}
\end{equation}

Substituting Equations~\ref{eqn:white-2} and \ref{eqn:white-5} into
Equation~\ref{eqn:white-3}, we find the following
expression for the electron volume number density energy distribution in the
nonrelativistic limit:
\begin{equation}
{{d^2 {\cal N}(E)} \over {dE\,dV}}\ =\ 3.04\,\times\,10^{24}\ {{A_0\ 
b({\gamma})} \over {E_{0,{\rm keV}}^{1.5}\,A_X}}\ \ 
\biggl({E \over E_0}\biggr)^{-({\gamma}+1.5)} \ \ \ \ {\rm electrons\
cm^{-3}\ keV^{-1}.}
\label{eqn:white-6}
\end{equation}
This formula indicates that for a given photon power-law
index $\gamma$, the electron energy power-law index
$\delta\,=\,\gamma+1.5$ at nonrelativistic energies in a thick-target model, i.e., the energy spectrum of the electrons entering the
target is actually steeper than the
resulting thick-target bremsstrahlung photon spectrum. This occurs
because the electron energy spectrum in the target is flatter than the
injected energy spectrum, due to the fact that electrons with low energy
lose their energy faster by collisions with ambient electrons, and thus are 
depleted in the target faster, than high-energy electrons.

The ultrarelativistic limit is somewhat complicated and simple expressions 
are not available\index{bremsstrahlung!ultrarelativistic case}. 
In terms of the spectrum, setting $v\,=\,c$ in
Equation~\ref{eqn:white-3} means that no additional power of $E$ is added when 
the electron flux is
converted to a number density, i.e., if this were the only change in the
ultrarelativistic limit we would find ${{d^2 {\cal N}(E)} \over {dE\,dV}}\
\propto\ E^{-({\gamma}+1)}$. In terms of the photon spectrum, this would
actually imply that for a pure electron energy power law $\delta$ the
bremsstrahlung photon spectrum would steepen in the ultrarelativistic
limit compared to the nonrelativistic limit by 0.5 in the spectral index,
i.e., $\gamma\,=\,\delta-1$ instead of $\gamma\,=\,\delta-1.5$. 
However, the electron-ion cross-section changes form in the
ultrarelativistic limit, and electron-electron collisions play a 
more important role in producing bremsstrahlung at energies above the 
electron rest mass\index{bremsstrahlung!electron-electron}.
Electron-electron bremsstrahlung
results in a photon spectrum shallower by~1 in the spectral index 
than electron-ion bremsstrahlung \citep{Ves88,KBE10}.
(The main energy loss of the incident electrons occurs
in $e-e$ collisions, but at lower energies HXR photons are predominantly 
produced by the large deflections in $e-p$ collisions.) 
The combined result of all these effects seems to be to flatten
the photon spectrum above 500 keV by about 0.5, i.e., $\gamma\,\approx\,\delta-2$.
\citet{McP90b} state
that a rough integration of the product of electron flux and
cross-section gives a flattening of the photon spectrum above 500~keV 
by $\log(E_\gamma/m_e c^2)$, although
their calculation refers to fluxes rather than volume number densities.
In any case, it does not seem to be readily feasible to derive
quasi-analytical approximations in the relativistic limit: 
a crude approach is to join the nonrelativistic limit smoothly onto a
power law~0.5 flatter above 500~keV.

\subsection{Thin-target limit}
\index{thin target}
\index{bremsstrahlung!thin-target}

In contrast to the thick-target limit, which is generally believed to apply
to footpoint HXR sources, in the thin-target limit\index{hard X-rays!thin-target} the electron energy
distribution evolves only slowly under the influence of collisions. 
The rate of bremsstrahlung emission depends on the ambient ion number density $n_i$
(\pccm) in the source (assumed to be much larger than the density of any
accelerated ions present).  
In this case, for a given photon spectrum (Equation~\ref{eqn:white-1}), we find that the
nonthermal electron energy distribution is

\begin{equation}
{{d^2 {\cal N}(E)} \over {dE\,dV}}\ =\ 
7.9\,\times\,10^{41}\ {{A_0\ C(\gamma)} \over n_{i}\ V_X\ E_{0,{\rm
keV}}^{0.5}} \ \biggl({E \over
E_0}\biggr)^{-(\gamma-0.5)} \ \ \ \ {\rm electrons\ cm^{-3}\ keV^{-1}},
\end{equation}

\noindent where $C(\gamma)\,=\,(\gamma\,-\,1)/B(\gamma\,-\,1,0.5)$ and
$V_X$ is the physical volume of the hard X-ray source. In the thin-target limit, $\delta\,=\,\gamma\,-\,0.5$, i.e., the electron
energy distribution is flatter than the photon energy spectrum by about~0.5 in the index, whereas in
the thick-target case the electron energy distribution is steeper than the
observed photon energy spectrum.

\subsection{Simplified gyrosynchrotron formulae}
\index{spectrum!gyrosynchrotron!simplified formulae}

Exact calculations of the gyromagnetic emission by an energetic electron
distribution
are somewhat complicated and do not lend themselves to simple analytic
formulae \citep[e.g., ][]{Ram69,Tak72}. Faster numerical approaches have
also been developed \citep[e.g.,][]{Pet81,Kle87,FlK10}. In order to have
rough analytic estimates of 
the number density of the radio-emitting electrons, we use
the Dulk \& Marsh (1982) approximation formulae \citep[see][for corrected versions]{Dul85}
for gyrosynchrotron emission.
\nocite{DuM82}
These describe a power-law distribution of mildly
relativistic electrons, and are valid for isotropic electron pitch-angle
distributions, harmonics in the range 20-100 and viewing angles
(between the line of sight and the magnetic field direction) in the
range 30\degr-80\degr.
The physical property of the radio source that is easiest to measure and least
dependent on a specific geometrical model is the
brightness temperature\index{brightness temperature}\index{radio emission!brightness temperature} at a point in the image, obtained at the highest
frequency and spatial resolution possible. 
Other quantities, such as
fluxes, require source areas to be known, and these are much more
model-dependent than a localized brightness temperature.
On the other hand, measuring the true brightness temperatures requires
excellent high-spatial-resolution imaging.
It is assumed that the radio spectral index ${\alpha_r}$\index{radio emission!diagnostics!spectral index} can be measured
from two optically-thin frequencies (i.e., two frequencies above the
flux peak in the radio spectrum). The expression for radio
brightness temperature at a frequency~$f$ produced by an electron
number density energy distribution 

\begin{equation}
{{d^2 {\cal N}(E)} \over {dE\,dV}}\ =\ {\cal N}_r\ {(\delta\,-\,1) \over E_r}\ \
\biggl({E \over E_r}\biggr)^{-\delta}\ \ \ \ {\rm electrons\ cm^{-3}\
keV^{-1}}
\label{eqn:white-8}
\end{equation}

\noindent (here ${\cal N}_r$ is the number of electrons per unit volume 
in the distribution above
the fiducial electron energy $E_r$, taken to be 10~keV by Dulk \& Marsh 1982) 
in a magnetic field of strength $B$ Gauss is 

\begin{equation}
T_B\ =\ \kappa\,T_{eff}\ =\ 3.1\,\times\,10^{-0.53\delta}\
(\sin \theta)^{-0.45+0.66\delta}\ \biggl({f \over
f_B}\biggr)^{-0.80-0.90\delta}\ \ {{\cal N}_r\,L \over B},
\label{eqn:white-9}
\end{equation}

\noindent where $\theta$ is the angle between the magnetic field and the
line of sight, $L$ is the line-of-sight depth through the source at
the point where $T_B$ is measured, and $f_B\ =\ 2.8\,\times\,10^6\,B$ Hz
is the electron gyrofrequency.\index{frequency!Larmor} 
Throughout this review we will use cgs units
except for photon and electron energies, where in solar physics it is 
traditional to use keV rather than ergs as the energy unit. 

Equation~(\ref{eqn:white-9}) results from multiplying
together the Dulk (1985) expressions for opacity $\kappa$ and effective
temperature $T_{eff}$. 
We proceed to 
use Equation~(\ref{eqn:white-9}) for $T_B$ to
determine the parameter ${\cal N}_r$ that appears in Equation~\ref{eqn:white-8}; note that ${\cal N}$ can also
be derived directly from the X-ray Equation~\ref{eqn:white-6} by equating it to
Equation~\ref{eqn:white-8}, as presented below. 

For convenience we write
\begin{equation}
{f \over f_B}\ =\ {{10^9\,f_{\rm GHz}} \over {2.8\,\times\,10^6\,B}}\ =\ 357
\ {f_{\rm GHz} \over B}
\label{eqn:white-10}
\end{equation}
\index{frequency!cyclotron}

\noindent (where $B$ is everywhere in units of Gauss) and find
\begin{equation}
T_B\ =\ 2.81\,\times\,10^{-2.00-2.83\delta}\
(\sin \theta)^{-0.45+0.66\delta}\ f_{\rm GHz}^{-0.80-0.90\delta}\ 
{\cal N}_r\,L\,B^{-0.20+0.90\delta}.
\label{eqn:white-11}
\end{equation}

\noindent Inverting this expression to obtain ${\cal N}_r$ and substituting into
Equation~\ref{eqn:white-8} we find
\begin{equation}
{{d^2 {\cal N}(E)} \over {dE\,dV}}\ =\ 35.3\ (\delta-1)\ 10^{3.83\delta-1}\
{T_B \over {L\,B^{-0.20+0.90\delta}}}\ (\sin \theta)^{0.45-0.66\delta}\ 
f_{\rm GHz}^{0.80+0.90\delta}\ E_{keV}^{-\delta},
\label{eqn:white-12}
\index{radio emission!microwave peak frequency!formula}
\end{equation}
in units of ${\rm electrons\ cm^{-3}\ keV^{-1}}$. The quantity
$T_B$ can be measured directly if imaging observations of sufficient 
resolution are available; $\theta$ is usually assumed to be close to 90$^\circ$
since this maximizes the gyrosynchrotron emissivity; $L$ can be
estimated from the radio images; and since ${\alpha_r}\ =\ 1.20-0.90\delta$ is 
the radio flux spectral index, the value of $\delta$ obtained from the hard
X-ray spectrum can be independently checked from 
the high frequency radio spectrum. The physical quantity in this
equation that cannot be derived from observations 
is $B$, which appears as a very high power. In principle this
too can be determined using the location of the radio peak frequency,\index{radio emission!diagnostics!microwave peak frequency}\index{radio emission!diagnostics!microwave peak frequency}\index{frequency!microwave peak}
which determines $B$ via the following gyrosynchrotron approximation:
\begin{equation}
f_{peak}\ =\ 2.72\,\times\,10^{3.00+0.27\delta}\
(\sin \theta)^{0.41+0.03\delta}\ ({\cal N}_r\,L)^{0.32-0.03\delta}\ 
B^{0.68+0.03\delta}.
\label{eqn:white-13}
\end{equation}

\noindent A peak frequency of 10~GHz typically corresponds to $B\ 
\approx$~600~G (roughly the 6th harmonic of the gyrofrequency), 
with only a weak dependence on other physical parameters
in the source \citep[e.g.,][]{DuM82}.
\index{spectrum!microwave!peak frequency}\index{frequency!microwave peak}

\begin{figure}
\centerline{\includegraphics[width=0.98\textwidth]{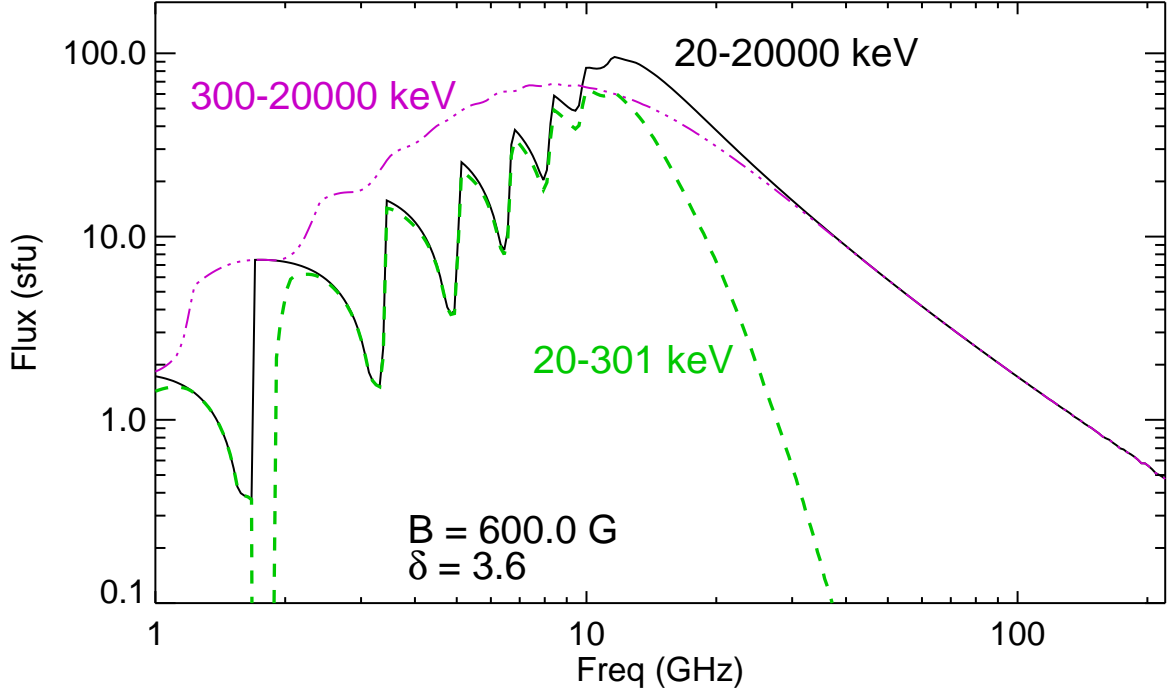}}
\caption[]{A comparison of gyrosynchrotron radio spectra from electrons
with different energy ranges. The black line is the spectrum produced by 
an electron energy power law from 20 to 20000 keV with a spectral index
of 3.6 in a constant magnetic field of 600 G. The green dashed line 
is the spectrum produced by  
the lower-energy electrons from 20 to 300 keV, and the purple dash-dot
line is the spectrum produced by the 300 to 20000 keV electrons. 
The calculation assumes a homogeneous loop of length 10\arcsec, width
3\arcsec\ and depth 1\arcsec, and an 
isotropic pitch-angle distribution for the radiating electrons. Radiative
transfer is taken into account in the calculations.
}
\label{fig:white-0}
\index{spectrum!microwave!illustration}
\end{figure}

Equations~(\ref{eqn:white-6}) and (\ref{eqn:white-12}) are two separate 
expressions for a volume density energy distribution and can be compared. When 
\index{electrons!spectra!HXR/radio comparison}
the spectral indices disagree by a large amount, as is often the case (with 
the energy distribution derived from radio data usually being flatter than that
derived from HXR data), then it is
usually assumed that they refer to different energy ranges: the
optically-thin microwaves are more sensitive to electrons above 300 keV, while
hard X-rays are usually dominated by electrons below 300 keV. This is
illustrated in Figure \ref{fig:white-0}, which compares the
gyrosynchrotron spectra of high- and low-energy electrons: electrons below
300 keV produce little radio emission above the spectral peak near 10~GHz (in the case of Figure~\ref{fig:white-0}, at 30 GHz, the 20-300 keV
emission is already two orders of magnitude weaker than the emission from
the full energy range). When the electrons below 100 keV are
present, they are sufficient to make the source optically thick at
frequencies below the spectral peak, and the harmonic structure is clear
(but will only be observed if the magnetic field is constant throughout
the source, which is unlikely to be the case); when
only electrons above 300 keV are present, they produce a smoother
spectrum
below the spectral peak because their harmonics are intrinsically 
\index{synchrotron emission}
broader.  The spectrum shows some flattening above 40~GHz in the regime
where relativistic particles dominate the emission.  Note that
the Dulk \& Marsh formulae were derived from a full relativistic calculation 
for the gyrosynchrotron emission, so there is no contradiction in comparing the
nonrelativistic thick-target bremsstrahlung results dominated by $<$500 keV
electrons with the relativistic gyrosynchrotron emission dominated by $>$500
keV electrons, as long as the underlying electron energy spectrum has an
unbroken power law (Equation~\ref{eqn:white-2}) from nonrelativistic  
to relativistic energies.

The discussion in this section relies on the paradigm that the
nonthermal electrons are accelerated in the corona and then propagate
down to the footpoints of the loops.\index{paradigms!precipitation from corona}
Another complication for the comparison between radio and HXR results 
that needs to be borne in mind is
that the number density appearing in Equation~\ref{eqn:white-6} refers to the 
density entering the thick target at the footpoints, i.e., at some 
depth in the chromosphere
depending on the electron energy \citep[e.g., ][]{BAK02}, 
while the radio emission comes from the corona and the number density
therefore refers to a different location in the loop. Implicit is also the
assumption that the pitch-angle distribution is everywhere isotropic,
corresponding to the fast pitch-angle scattering limit; an anisotropy in
the pitch-angle distribution will affect the comparison (discussed
further below).\index{scattering!pitch-angle}

\subsection{Radio flux from a thick-target hard X-ray spectrum}
\index{thick-target model!radio flux}

We may now derive an expression for the radio flux from an
integrated hard X-ray spectrum under the thick-target assumption. 
Given the radio brightness temperature, 
in this case the Dulk and Marsh approximation (Equation~\ref{eqn:white-9}), we obtain the total
radio flux from the standard expression \citep[e.g.,][equation 14]{Dul85}:
\begin{equation}
S\ =\ {{k_B\,f^2 } \over {c^2}} \ \int\ T_B \ d\Omega,
\label{eqn:white-17}
\end{equation}
where the integral is over the solid angle $\Omega$ subtended
by the radio source on the sky. We may replace the solid angle element
$d\Omega$ with $dA/D^2$, where $A$ is the actual physical area of the
source on the sky and $D$ is the distance of the source (in this case,
$D$ is one astronomical unit, 1.5 $\times$ 10$^{13}$ cm). 
From Equation~\ref{eqn:white-9} we may write
\begin{equation}
S\ =\ {{k_B\,f^2 } \over {c^2}} \ \int\ 
2.81\,\times\,10^{-2.00-2.83\delta}\
(\sin \theta)^{-0.45+0.66\delta}\ f_{\rm GHz}^{-0.80-0.90\delta}\ 
\,B^{-0.20+0.90\delta} \ {\cal N}_r\ {{L\,dA} \over {D^2}}.
\label{eqn:white-18}
\end{equation}
Now we note that $\int\-{\cal N}_r\ L\,dA$ is the integral of the
number density per unit volume (above the reference energy $E_r$)
over the volume of the source, i.e., it
is the total number of electrons in the emitting volume, ${\cal N}_{tot}$.
Combining
the physical constants (note that one solar flux unit, or ``sfu,'' is 10$^{-19}$
ergs cm$^{-2}$ s$^{-1}$ Hz$^{-1}$ in cgs units) and setting the angular 
factor to unity as an
approximation (e.g., for $\theta$ = 70\degr\ and $\delta$ = 3.0, 
$(\sin \theta)^{-0.45+0.66\delta}\,=\,0.9$), we find
\begin{equation}
S_{sfu}\ =\ 1.9\,\times\,10^{-28.0-2.83\delta}\ {\cal N}_{tot}\ 
\ f_{\rm GHz}^{1.20-0.90\delta}\ \ B^{-0.20+0.90\delta}. 
\label{eqn:white-19}
\end{equation}

\subsection{The effect of anisotropy}
\index{radio emission!gyrosynchrotron!anisotropy}

The discussion presented above assumes an isotropic pitch-angle
distribution for the radiating electrons.
\index{electrons!pitch-angle distribution}
This is appropriate, e.g.,
when strong pitch-angle scattering in the flare loops, by collisions or
by wave-particle interactions, isotropizes the particle distribution.\index{wave-particle interactions}\index{electrons!anisotropy}\index{scattering!pitch-angle}
This assumption is necessary to derive simple analytic formulae that can
be applied \citep[weak anisotropy is included in the papers
by][]{Pet82,Rob85,Kle87}, but may not be appropriate in many cases. 
\citet{FlM03a} discuss the effect of pitch-angle anisotropies on 
gyrosynchrotron emission in some detail. 
A single relativistic electron emits most of its radiation in
a narrow cone around the direction of motion, but particles with small
pitch angles barely radiate at all, so there can be large differences in
the emission of electrons with large and small pitch angles. Tied to
this question is the evolution of the pitch-angle distribution of a
population of magnetically-trapped electrons both in time and in space.
In a loop with significant variation in magnetic field strength between
the footpoints and the loop top, conservation of the first adiabatic
invariant\index{transport!adiabatic invariants!pitch-angle distributions}\index{electrons!adiabatic invariants}
means that the pitch angle of an electron will increase as it
\index{magnetic structures!mirror geometry}
propagates towards stronger magnetic fields, and the electron will
reflect at the height where its pitch angle reaches 90\degr\ (``magnetic
mirroring'').\index{accelerated particles!adiabatic invariants}\index{adiabatic invariants}
Electrons with pitch angles sufficiently small for them to 
reach the solar chromosphere will ``precipitate'' (i.e., lose their
energy immediately in collisions and stop) there and be lost from the
corona. When gradual pitch-angle scattering takes place, without renewed
injection of electrons, one ends up with a population of electrons
with large pitch angles trapped near the loop top.\index{precipitation}\index{scattering!pitch-angle}

\citet{FlM03a} \citep[with corrections in ][]{FlM03b} investigate changes
to the gyrosynchrotron spectrum due to various forms of pitch-angle 
anisotropy at different observer viewing angles. Changes are more
pronounced for viewing angles close to the magnetic field direction
rather than perpendicular to the magnetic field direction in the source.
\index{polarization!gyrosynchrotron}
The degree of polarization tends to increase as the anisotropy of the
pitch-angle distribution strengthens, and the spectral index in the 
optically-thin limit is larger (steeper spectra) than for the isotropic
pitch-angle distribution.
In particular, loss-cone distributions viewed along the magnetic field
\index{electrons!distribution function!loss-cone}
show much weaker emission than isotropic distributions, and steeper
high-frequency (i.e., optically thin) spectra: this can explain some
observed differences between loop ends and loop tops for disk flares
where the legs of the loop are viewed along the magnetic field and the
loop top is viewed orthogonal to the magnetic field. 

\subsection{Transport effects}
\index{radio emission!transport effects}

Particle transport effects can also influence the appearance of
microwave and hard X-ray sources, and delays between the two types of
emission.  Simultaneous imaging observations of radio and HXR sources can
place physical constraints (loop lengths, source separations) on the
flare site. We will not address these issues in detail here. We note
that \citet{MRG06}
discuss the ways in which both pitch-angle distributions and particle
transport effects can affect the microwave appearance of flaring loops\index{models!trap-plus-precipitation}.
\citet{MYM08} carry out detailed calculations of a
trap-plus-precipitation model for a flare observed by \textit{\textit{RHESSI}} and NoRH.
This flare exhibited HXR emission from two footpoints and a loop-top
microwave source, and the authors argue that their model can explain the
observation that the radio-emitting electrons appear to have a spectrum
harder than that of the HXR-emitting electrons. 
\index{looptop sources!microwaves}
As time proceeds,
higher energy electrons will tend to concentrate in the trapped
population of coronal loops, and this can explain observations such as
the gradual hardening of the microwave spectrum even when the HXR
spectrum is softening \citep[e.g., ][]{Nin07a,Nin08b}.

\section{Thermal bremsstrahlung radio emission from flares}\label{sec:white-3}
\index{radio emission!thermal}\index{free-free emission}

For completeness we include a brief discussion of thermal radio
emission. The opacity due to bremsstrahlung 
of a uniform plasma of typical solar composition 
at temperature $T_e$ and density $n$ can be approximated by 
\citep[e.g.,][]{Dul85}
\begin{equation}
\kappa\ =\ 0.2\,{{n^2} \over {T_e^{1.5}\,f^2}}\ \ \ {\rm cm}^{-1},
\label{eqn:white-20}
\end{equation}
where $n$ is measured in cm$^{-3}$, $T_e$ in Kelvin and
frequency $f$ in Hz. From Equation~\ref{eqn:white-17}, the corresponding radio flux 
is given by 
\begin{equation}
S\ =\ k_B/c^2\,\Omega\ \left\{ \begin{array}{ll}
T_e\,f^2 & \ \ \ \ \ {\rm optically\ thick:}\ \kappa L \gg\ 1 \\
{0.2\,n^2\,L / T_e^{0.5}} & \ \ \ \ \ {\rm optically\ thin:}\ \ \ \kappa L \ll\ 1,
\end{array} \right.
\label{eqn:white-21}
\end{equation}
where $L$ is the thickness of the source along the line of
sight (in units of cm) and $\Omega$ is the projected solid angle on the sky 
occupied by the source. As above, the radio flux is traditionally
quoted in units of sfu. As a crude rule of thumb, we may write
Equation~\ref{eqn:white-21} as
\begin{equation}
S\ =\ 2\ \ \biggl({T_B \over {10^6 \,{\rm K}}}\biggr)\ \biggl({f \over {\rm 10\ GHz}}\biggr)^2
      \ \biggl({r \over {10^{\prime\prime}}}\biggr)^2 \ \ {\rm sfu},
\label{eqn:white-22}
\end{equation}
where $r$ is the radius of the area in which the brightness
temperature is $T_B$ (with $T_B\,=\,T_e$ in the optically-thick limit
and $T_B\,=\,\kappa L T_e$ in the optically-thin limit).

Generally in solar flares, we find that the heated
coronal plasma is optically thin to thermal bremsstrahlung 
above about 10 GHz\index{radio emission!thermal!optical depth}, 
and this can be verified because, by
Equation~\ref{eqn:white-21},
the flux spectrum should be almost independent of frequency (or
equivalently, the brightness temperature spectrum, obtained from 
spatially-resolved images, varies as $f^{-2}$). 
A \textit{GOES} class M1~flare typically produces a peak thermal flux of 1-5~sfu 
in the optically-thin flat-spectrum limit
from the 10$^7$ K plasma to which \textit{GOES} is sensitive,
while an X1 flare produces of order 10-50 sfu of thermal emission\index{GOES@\textit{GOES}!and radio fluxes}\index{free-free emission!microwaves}\index{free-free emission!optically thick}.
At low, optically thick, frequencies the spectrum rises as $f^2$.
The $n^2/T^{0.5}$ dependence in the optically thin limit means that thermal
bremsstrahlung is strongest in cool dense plasmas. In fact, if we take a
given volume of optically thin plasma and heat it up without changing the
density, the radio flux from the plasma will actually decrease due to the
$T^{-1.5}$ dependence of the opacity. The \textit{GOES} soft X-ray data are
most sensitive to plasma at a temperature of order 10 MK, and if any
cooler plasma is also present then the thermal radio emission will be
larger than the flux level inferred from the \textit{GOES} data alone. It should
also be noted that the radio emission is produced by electrons whereas
the soft X-ray emission has a large contribution from lines of highly-ionized
atoms of elements heavier than hydrogen, so quantitative comparison of the 
two also depends on the abundance distribution of the radiating plasma
\citep[see the discussion in][]{WTS05}.

\section{Physical properties of flare-accelerated particles from hard X-rays 
and microwaves}\label{sec:white-4}

\begin{figure}
\centerline{\includegraphics[width=0.98\textwidth]{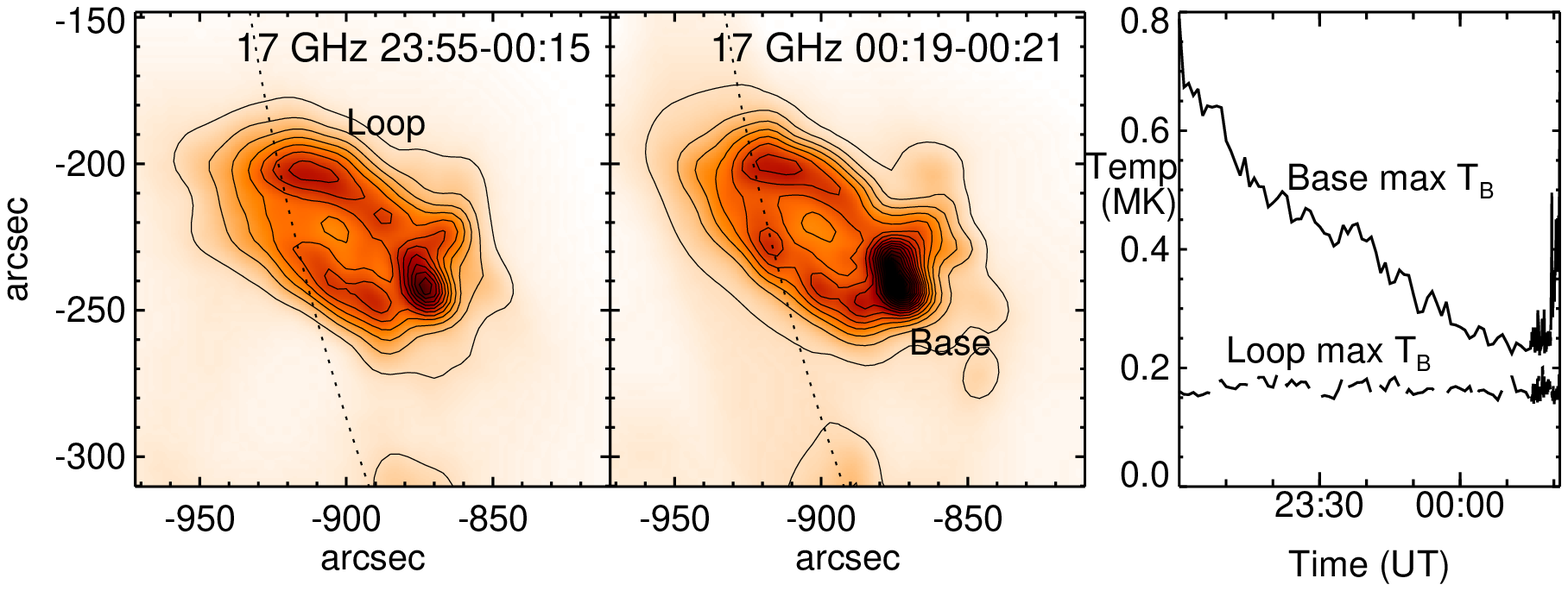}}
\caption[]{The evolution of pre-flare emission at 17~GHz in SOL2002-07-23T00:35 (X4.8). 
\index{flare (individual)!SOL2002-07-23T00:35 (X4.8)!radio observations}
The left panel shows an image averaged over the period 20~minutes 
prior to the impulsive phase, showing a distended loop with a
compact source at its base. The central panel shows the image at 00:20~UT, with the source at the base brightening significantly. The contour
interval is 2 $\times$ 10$^4$ K in both images. The right
panel shows the light curves of the maximum 17~GHz brightness 
temperature in the compact source at the base (solid curve) and in the rest of
the loop (dashed line).
}
\label{fig:white-020723-pre-flare}
\end{figure}
\index{flare (individual)!SOL2002-07-23T00:35 (X4.8)!illustration}

As discussed in Section~\ref{sec:white-2}, if electrons with the same
energy distribution produce both the microwave and hard X-ray emission
from a solar flare, there should be straightforward relationships
between the spectral indices of the two emissions. 
However, there has
been a long record of comparisons finding that in fact the radio
and hard X-ray spectral indices are not compatible with the same
electron energy distribution \citep[e.g.,][]{KWG94,SWG00}.
\index{electrons!radio/HXR spectral mismatch}
Since the
typical (initial) energy of an electron emitting a HXR photon of energy 
${\cal E}$ is $1.5{\cal E}-3{\cal E}$, with higher electron energies for
flatter energy spectra \citep[e.g.,][]{KDK88,AsS96a}, 
and microwave emission at higher frequencies 
is typically dominated by electrons with energies above 300~keV, it is
desirable to have the hard X-ray spectral index at energies up to
several hundred keV when making this comparison.
Previously, such comparisons between microwave and
HXR/$\gamma$-ray photon spectra up to several hundreds of keV have been possible with
\textit{GRANAT}/PHEBUS data \citep{TVB98,VTB99}.\index{satellites!GRANAT@\textit{GRANAT}} 
Those studies found consistency between the spectral indices of the
radio-emitting electrons and the HXR/$\gamma$-ray-emitting electrons at
energies above several hundred keV.

\begin{figure}
\centerline{\includegraphics[width=0.85\textwidth]{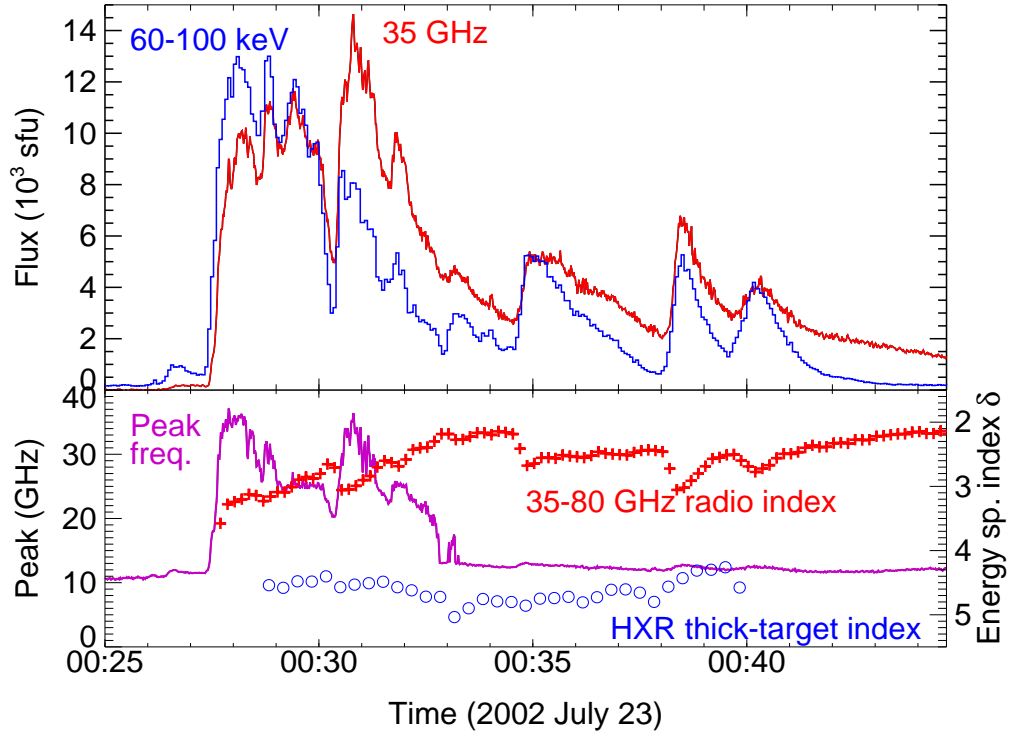}}
\caption[]{Comparison (upper panel) of the \textit{RHESSI} 60-100 keV hard X-ray
light curve (blue histogram) and the NoRP 35 GHz light curve (red solid curve)
for the well-observed flare SOL2002-07-23T00:35 (X4.8), as well as (lower panel)
the time evolution of the radio spectral peak frequency (purple solid line) and
the radio spectral index from 35 to 80~GHz converted to an electron energy
spectral index assuming gyrosynchrotron emission from an optically-thin
source (red plus symbols, uncertainty $\pm$0.3).
For comparison, the thick-target electron energy index obtained from
the \textit{RHESSI} 100-400 keV spectrum is also shown (blue open circles; formal
uncertainty $\pm$0.2). From \citet{WKS03}.
}
\label{fig:white-020723}
\index{flare (individual)!SOL2002-07-23T00:35 (X4.8)!illustration}
\index{frequency!microwave peak!illustration}
\end{figure}

\subsection{SOL2002-07-23T00:35 (X4.8)}
\index{flare (individual)!SOL2002-07-23T00:35 (X4.8)!radio properties}

The first \textit{RHESSI} flare for which this was possible was SOL2002-07-23T00:35 (X4.8).  
Before the main phase of this event, a remarkable
distended precursor radio loop is seen,
shown in Figure \ref{fig:white-020723-pre-flare} \citep{ANS06}.
The elongated loop structure is visible in both the 17~and 34~GHz images,
extending over 80\arcsec\ to the north-east from the subsequent site of
the flare (located at the base of this loop, marked in the middle panel
of Figure \ref{fig:white-020723-pre-flare}).
\index{flare (individual)!SOL2002-07-23T00:35 (X4.8)!radio observations}
In the 75 minutes for which NoRH images are available prior to the flare,
the loop does not seem to change
shape, nor are there significant changes in its brightness. On the other
hand, the compact source at the base of the loop, which is not visible
at 34 GHz and therefore may be nonthermal, 
declines steadily over this period until 00:18~UT, when
pre-flare activity commences. EUV images of this region show no feature
corresponding to the radio loop: it is difficult to measure its
spectrum, but it may be thermal, suggesting that it is material much
hotter than the $\sim$10$^6$~K range to which the EUV images are
sensitive.

Once the flare starts, intense nonthermal gyrosynchrotron emission from
the main flare loops takes up the dynamic range of the images and the pre-flare loop is no longer visible.
\citet{WKS03} carried out a careful study of the main flare 
using radio imaging and spectral data from NoRH together with \textit{\textit{RHESSI}}
data. 
In common with other large flares, this event
had a high turnover frequency\index{radio emission!gyrosynchrotron!turnover frequency}\index{frequency!microwave peak}
in its radio spectrum, necessitating the
use of radio fluxes above 30~GHz in order to estimate the radio spectral
index on the optically-thin high-frequency side. The imaging data were
consistent with the picture in which the high-energy HXRs were
emitted from the footpoints of coronal loops visible in the radio images. 

Radio and HXR light curves for this event 
and the results of fits to the radio and HXR
spectra are shown in Figure \ref{fig:white-020723}. 
The structure of the 35~GHz light curve is representative of all
radio frequencies above 4~GHz, and 60-100~keV is representative of
the light curves for all hard X-rays above 30~keV; below 30~keV, the
X-rays show less temporal fine structure in their light curves \citep{HSS03}.
The impulsive phase consists of a number of spikes and dips that are
all seen in both the radio and HXR light curves.
Different peaks have relatively different heights at
the two wavelengths, but over the $\sim$20 minutes of the impulsive
phase the similarity in time profiles is striking, including the brief
sharp dip at 00:30:20 UT. 
Note that the rapid drop in radio emission\index{trapping!time scale}
exhibited during this dip implies that the trapping time for electrons
in the corona must be very short (of order seconds) in this event.

The lower panel in Figure~\ref{fig:white-020723} represents the spectral 
evolution of the radio
emission using two parameters: the peak frequency (solid line), and the radio
spectral index from 35 to 80~GHz (plus symbols), converted to the equivalent 
electron energy spectral index. The peak frequency in the spectrum,
i.e., the frequency at which the radio flux is largest at any time,
represents the boundary between the lower frequency emission from the
regime where the source is optically thick and higher frequency emission from 
the regime where the source is optically thin.  For comparison, the HXR
spectral index from 100-400 keV, converted to
an electron energy spectral index by assuming thick-target emission,
is also plotted (as circles) in the lower panel, and it is clear that the two 
energy spectral indices are inconsistent with one another: as usual
\citep[e.g.,][]{KWG94,SWG00}, the radio
data indicate a much flatter energy spectrum than do the HXR data.
\index{electrons!spectra!HXR/radio comparison}

Applying the quantitative analysis above to the hard X-ray data, we can
estimate electron number densities in this event. The 40-400 keV photon
energy spectrum at 00:35:00~UT,
when the radio spectral peak is well below 34~GHz, has the
form $9\,(E_\gamma/{\rm 50\,keV})^{-3.2}$ photons cm$^{-2}$ keV$^{-1}$
s$^{-1}$. The corresponding flux of electrons striking the chromosphere
is $2\,\times\,10^{35} (E/{\rm 20\,keV})^{-4.2}$ electrons
keV$^{-1}$ s$^{-1}$. To convert this to a number density we need to
divide by the footpoint area $A$. The \hsi\ images do not resolve the
footpoints at a resolution of 7\arcsec\ and images made using grid 1
suggest a size as small as 2\arcsec. If we adopt
$A$ = 10$^{16}$ cm$^2$, corresponding to a diameter of 2\arcsec, 
using the formalism of the previous section we derive an energy
distribution for the electron number density at the footpoints of
$5\,\times\,10^{9}\ (E/{\rm 20\,keV})^{-4.7} \ (A/10^{16}\,{\rm cm}^2)
\ \ {\rm electrons\ cm^{-3}\ keV^{-1}}$\index{footpoints!nonthermal electron density}.

If we take the electron energy spectral index of $-4.7$ derived from HXR 
and apply it to the radio
brightness temperature as discussed in the previous section, then the
radio data also require very high densities.\footnote{Note that the fact
that the energy spectral index derived from HXR is steeper than that
derived from radio spectra implies that there are fewer high--energy
electrons in the spectrum, and since these are the electrons that
dominate the microwave emission, a larger total number of electrons is
then required.} Based on
gyrosynchrotron theory, the brightness
temperatures of over 10$^9$~K achieved at 17~GHz at the peak of the
flare can only be produced by an electron energy spectrum as steep as $-4.7$ 
if the harmonic number is in excess of~30
\citep[see figures in ][confirmed by accurate numerical calculations]{DuM82}.
Such high harmonic numbers imply a relatively small magnetic field
strength, no more than 200~G, and to achieve the same radio brightness
temperature with the lower
value of $B$, nonthermal densities over 10$^{10}$ cm$^{-3}$ above 20
keV are required, similar to those found above from the HXR data.
The low value for $B$ is consistent with the location of the brightest
radio emission at the top of the arcade of loops. \citet{CaL10} found
that the superhot component in this flare (temperature of order 30-40 MK)
required a magnetic field strength of at least 500~G to confine it,
consistent with a picture in which 
the bulk of the nonthermal particles are located at a
greater height than the superhot component. Even larger values of
nonthermal electron density, with much stronger magnetic fields, 
were deduced by \citet{RMK04} in the case of SOL2001-08-25T16:45 (X5.3).
\index{flare (individual)!SOL2001-08-25T16:45 (X5.3)!radio emission}

So, unfortunately, this extremely well observed event only adds to the
problem of reconciling energy spectra derived from radio and HXR data.
Since the radio and HXR profiles are so similar, we cannot argue
that the radio emission comes from a long-lived population of trapped
electrons while the hard X-rays come from directly precipitating
electrons, as occurs in other events where the time profiles clearly
differ \citep[e.g., ][]{RWK99}: the
radio-emitting electrons have the same time behavior as the hard
X-ray emitting electrons and should have a common origin and common
evolution. 
There is a flattening in the \hsi\ $\gamma$-ray spectra by 
about 0.5 in the index, but no sign in the data up to 8~MeV of the flattening by~2 implied 
by the radio data \citep{SSM03}, so we argue that the radio spectral
indices in Figure~\ref{fig:white-020723} are not compatible with 
the \hsi\ observations. We are
forced to assume that the high-frequency radio emission is dominated by
a region
with a high turnover frequency so that the 35-80~GHz spectrum does not
represent optically-thin emission, even in the later stages of the
flare, and the true optically-thin
radio spectral index is steeper than derived above.
\citet{NWS91} reached a similar conclusion for their event.
Both radio and HXR data require extreme number densities of
nonthermal electrons to be accelerated in the energy release: over
10$^{10}$ cm$^{-3}$ above 20~keV.\index{electrons!dominant tail population}
These numbers imply an acceleration mechanism of very high efficiency.
\index{acceleration!electrons!high efficiency}

\begin{figure}
\centerline{\includegraphics[width=\textwidth]{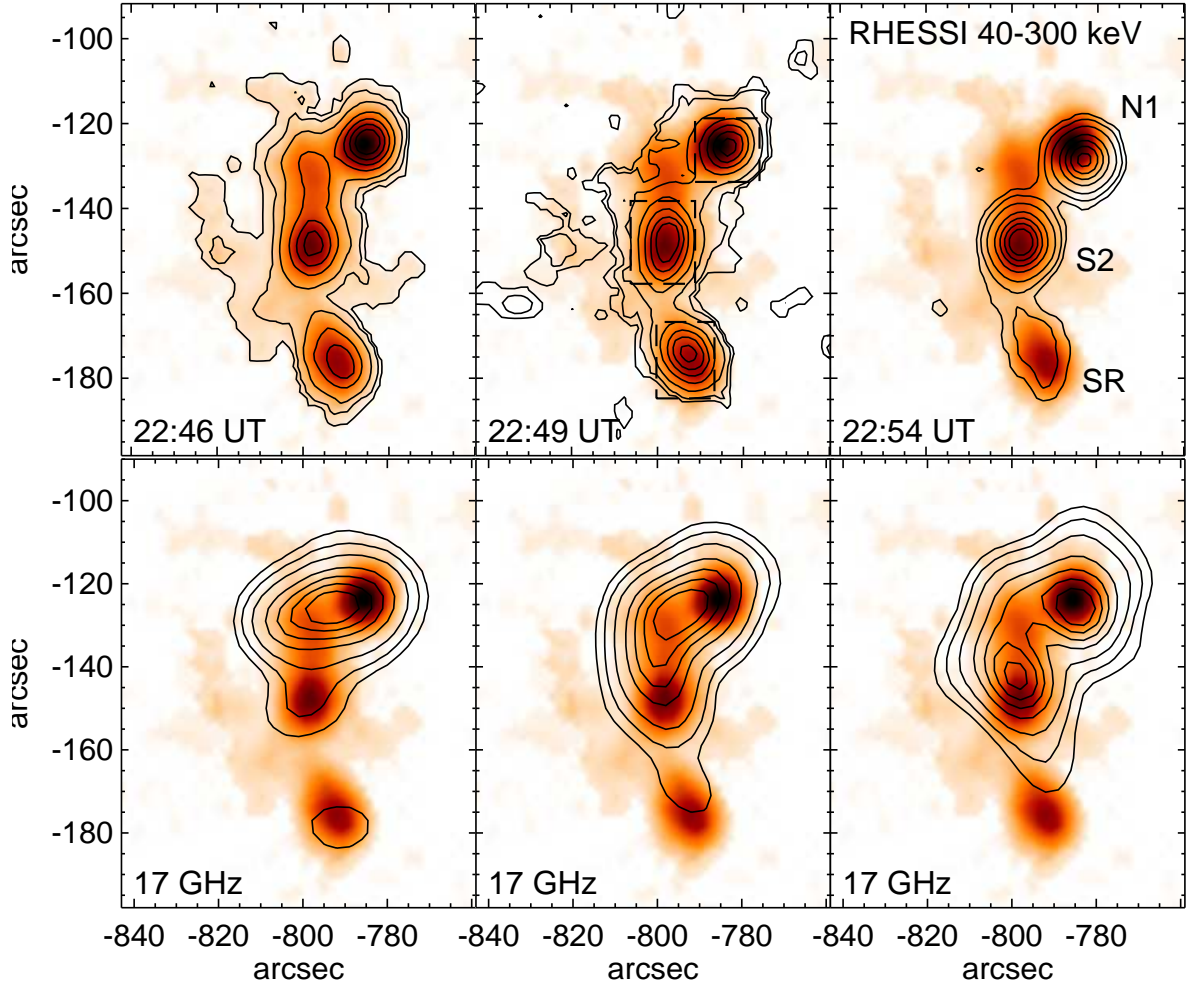}}
\caption{Upper panels: hard X-ray images at each of the three main peaks in the
HXR light curve (contours) of SOL2003-06-17T22:55 (M6.8). 
The images result from summing over
the 40\,-\,300 keV range. The background image in each panel is
the HXR image for 22:50 UT so that changes in morphology from one peak
to the next can be seen.  The resolution of the HXR images is
9\arcsec. The middle panel shows the regions used for the HXR
spectra of sources S2, N1, and SR as dotted boxes. Lower panels:
contours of simultaneous 
17 GHz emission plotted over the HXR image of the flare from
the upper left panel. Contours are at 4, 8, 16, 32, 48, 64, and
80\% of the peak in each image. From \citet{KGW09}.}
\label{fig:white-030617}
\end{figure}
\index{flare (individual)!SOL2003-06-17T22:55 (M6.8)!illustration}

\subsection{SOL2003-06-17T22:55 (M6.8)}
\index{flare (individual)!SOL2003-06-17T22:55 (M6.8)!radio properties}

Another well-observed flare is SOL2003-06-17T22:55 (M6.8) \citep{KGW09}.
\index{flare (individual)!SOL2003-06-17T22:55 (M6.8)!radio emission}
This flare is one of the few in
which emission up to energies exceeding 200~keV can be imaged in
hard X-rays, and we can investigate the HXR spectra of
individual sources up to this energy. The morphology of the event is
shown in Figure \ref{fig:white-030617}.\index{flare types!two-ribbon}\index{ribbons}\index{coronal sources}\index{hard X-rays!coronal sources}\index{flare (individual)!SOL2003-06-17T22:55 (M6.8)!radio emission}\index{flare (individual)!SOL2003-06-17T22:55 (M6.8)!coronal hard X-rays}
It takes the form of a filament eruption followed by a large
double-ribbon flare with nonthermal emission stretching along a
north-south axis over a neutral line.\index{filaments!eruptive}
Three peaks each separated by about 
5~minutes (times labeled in Figure~\ref{fig:white-030617}) are seen in the HXR 
light curve, and are matched in the radio data. In this case the radio and HXR
morphologies are very similar, with both showing sources seeming to
straddle the neutral line: the southernmost source is brightest during
the first peak and then fades relative to the other peaks. Note that HXR
and radio emission are both seen over a distance of more than 60~Mm
along the neutral line in this one event: nonthermal energy release must
have taken place throughout a large volume.

\begin{figure}[t]
\centerline{\includegraphics[width=\textwidth]{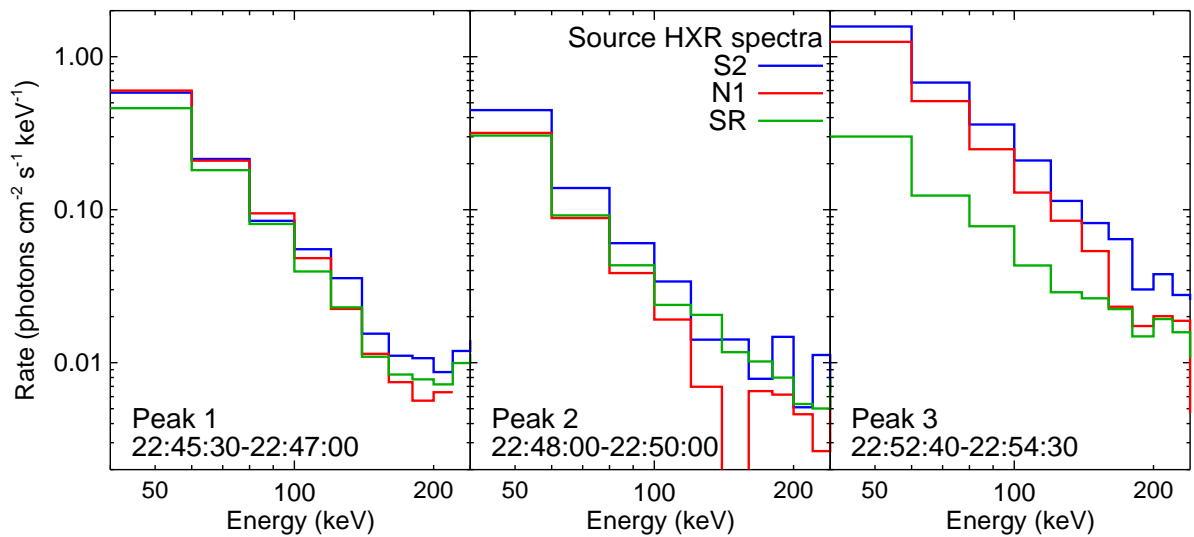}}
\caption{Hard X-ray spectra at each of the three main peaks in the
light curve for each of the three sources in the HXR images of SOL2003-06-17T22:55 (M6.8). 
These are derived from image cubes made in 20~keV channels from 40 to
300 keV. From \citet{KGW09}.
 } 
\label{fig:white-030617spec}
\index{flare (individual)!SOL2003-06-17T22:55 (M6.8)!illustration}
\end{figure}

The distinctive feature of this event is that we can clearly separate
three spatially distinct HXR sources during three different temporal
peaks, and compare the energy spectra in each case. Figure
\ref{fig:white-030617spec} shows the \hsi\ spectra for the three sources
outlined by boxes in the upper middle panel of
Figure~\ref{fig:white-030617} during each of the three peaks, and Table
\ref{tab:white-030617} reports the spectral indices determined from 
power-law fits to these spectra \citep{KGW09}.

\begin{table}[h]
\begin{tabular}{l|cccc}
\hline
Peak & S2 (middle) & N1 (north) & SR (south) & Spatially integrated \\
\hline
22:45:30-22:47:00 UT& 3.1 $\pm$ 0.3 & 3.3 $\pm$ 0.3 & 3.0 $\pm$ 0.4 &
3.3 (280) 2.5 $\pm$ 0.2 \\
22:48:00-22:50:00 UT& 3.3 $\pm$ 0.1 & 3.5 $\pm$ 0.1 & 2.8 $\pm$ 0.5 &
3.4 (210) 2.0 $\pm$ 0.2 \\
22:52:40-22:54:30 UT& 2.7 $\pm$ 0.1 & 2.9 $\pm$ 0.2 & 2.1 $\pm$ 0.6 &
2.5 (120) 3.4 $\pm$ 0.2 \\
\hline
\end{tabular}
\caption{Power-law fits to the photon spectral index $\gamma$ of
individual sources in each of the three main peaks in the SOL2003-06-17T22:55  (M6.8)
light curve derived from images in different energy bins,
together with the fit to background-subtracted 50\,-\,400 keV
spectra from the \textit{RHESSI} front detectors. For the spatially-integrated 
spectra, the numbers are the results of a broken
power-law fit: the spectral index at energies below the break, the
break energy (keV) in parentheses, and the spectral index above
the break. Uncertainties in the fits to the break energies are
typically large (tens of keV). }
\label{tab:white-030617}
\index{flare (individual)!SOL2003-06-17T22:55 (M6.8)!illustration}
\end{table}

The fits to the spatially-resolved spectra assume a single power
law over the 40\,-\,240 keV range, while the fits to the
integrated spectra assume a broken power law over the range
50\,-\,400 keV. For peaks 1 and 2 the spectral break in the power
law is above 200 keV and the fitted spectral index below 200~keV
generally matches the fits to the spatially resolved spectra,
which are dominated by photons below 120~keV, while the fit above
the break gives a flatter spectrum. The uncertainties in the
fits above the break are large due to the small number of 
high-energy photons detected, so one cannot conclude that there is
flattening of the electron energy spectrum at high energies during the
first two peaks (although note that fits to the \textit{RHESSI}
rear-detector spectra, more sensitive than the front detectors at high
photon energies, agree with the results in Table \ref{tab:white-030617}
to within the uncertainties).
)
We can also look at radio spectra for the individual peaks. NoRH images
can be used to measure the spectral index for the three individual
sources from 17~to 34~GHz, and again it is found that the inferred
energy spectra are flatter than the energy spectra inferred from the HXR
data with a difference in spectral index of order 1-2\index{electrons!spectra!HXR/radio comparison}\index{observatories!Nobeyama polarimeters (NoRP)}\index{observatories!Nobeyama polarimeters (NoRP)}.
Non-imaging data from the Nobeyama polarimeters (NoRP) up to 80~GHz suggest steeper radio
spectra above 34~GHz, but there is a large uncertainty in the
calibration of the NoRP 80~GHz data during this period and this leads to
large uncertainties in the high-frequency spectral index.
We do see significant anomalies between the radio and HXR properties of
the sources: source SR is relatively bright at
34 GHz until late in the event, suggesting a very flat energy spectrum;
the radio spectra do not show the same trend that all the spectra
are flatter during peak~3. 
However, again we have major uncertainties in
the location of the peak in the radio spectra of the individual sources
since we only have images at two frequencies, and therefore  we cannot
determine whether the spectral
index from 17~to 34~GHz actually represents the true optically-thin
spectral index: if the 17~GHz source is partially optically thick 
then this frequency range will show a spectrum flatter than the true 
optically-thin spectral index\index{radio emission!microwaves!difficulty of determining optical depth}. 
These problems again emphasize the need
to have routine data at more than one frequency above 40~GHz if we are to
be confident in measuring the true optically-thin radio spectral index
in large flares.

The general conclusion of Figure~\ref{fig:white-030617spec} and
Table~\ref{tab:white-030617} is that a given peak shows the same energy
spectrum (from HXRs) in all three spatial locations, but it may differ from
one peak to the next: peaks 1 and 2 clearly have steeper spectra
than peak 3 (below 120 keV) in all three sources and in the
integrated spectra. This suggests either that the electron
acceleration mechanism has the same physical characteristics over
a large spatial scale (5 $\times$ 10$^4$ km) or a more localized
accelerator distributes electrons over the full volume. The
challenge for the first interpretation is the fact that all
sources show a flattening of their spectra in peak 3 after being
steeper in peaks 1 and 2: how can sources so far apart have their
characteristics change in the same way? For example, if
acceleration is due to stochastic acceleration by wave turbulence\index{turbulence},
how is turbulence generated with identical properties over such a
large volume\index{acceleration!stochastic!radio observations}?
On the other hand, a localized accelerator that can
distribute nonthermal electrons over a distance of 
5~$\times$~10$^4$~km is difficult to reconcile with the usual picture of
post-flare loops in two-ribbon flares that are typically much
shorter than the ribbons and straddle the neutral line rather than
parallel the ribbons.

\subsection{Coronal hard X-ray sources}
\index{coronal sources}

The traditional picture of HXR arising from bremsstrahlung when
nonthermal electrons strike the dense chromosphere is consistent with
most observations. HXR sources in the corona are much less often observed, at
least in part due to instrumental issues: imaging HXR telescopes typically
have a limited dynamic range, and this limits their ability to see
fainter thin-target 
emission from the low-density corona if it is competing with very
bright thick-target footpoint emission from
the high-density chromosphere.  
Coronal HXR emission is almost certainly
present in most flares but difficult to observe, and a recent 
comprehensive review is devoted to this subject \citep[see][]{KBC08}. 

There is one circumstance in which coronal emission does not have to
compete with bright footpoint emission, and that is when the footpoints
of flaring loops are obscured from our view by the solar limb\index{coronal sources}\index{hard X-rays!coronal sources}\index{occulted sources}.
Such occulted flares have long been our main source of knowledge of coronal
emission \citep[e.g.,][]{FrD71,McK75,KAE79}, and this continues to be true 
in the \hsi\ era. \citet{KHG10} observed a flare on 2007 December~31 that
\index{flare (individual)!SOL2007-12-31T01:11 (C8.3)!radio observations}
occurred 12\degr\ behind the limb as seen from Earth, but happily the
event was within the field of view of telescopes on the \textit{STEREO-B}\index{satellites!STEREO@\textit{STEREO}}
spacecraft trailing the Earth in its orbit. \hsi\ observed remarkably strong
emission from this flare, up to 80~keV, despite the fact that the
footpoints were occulted from \hsi's view. The 50~keV photon flux was comparable
to what we see from the footpoints in a \textit{GOES} class~M flare.

\begin{figure}[t]
\centerline{\includegraphics[width=\textwidth]{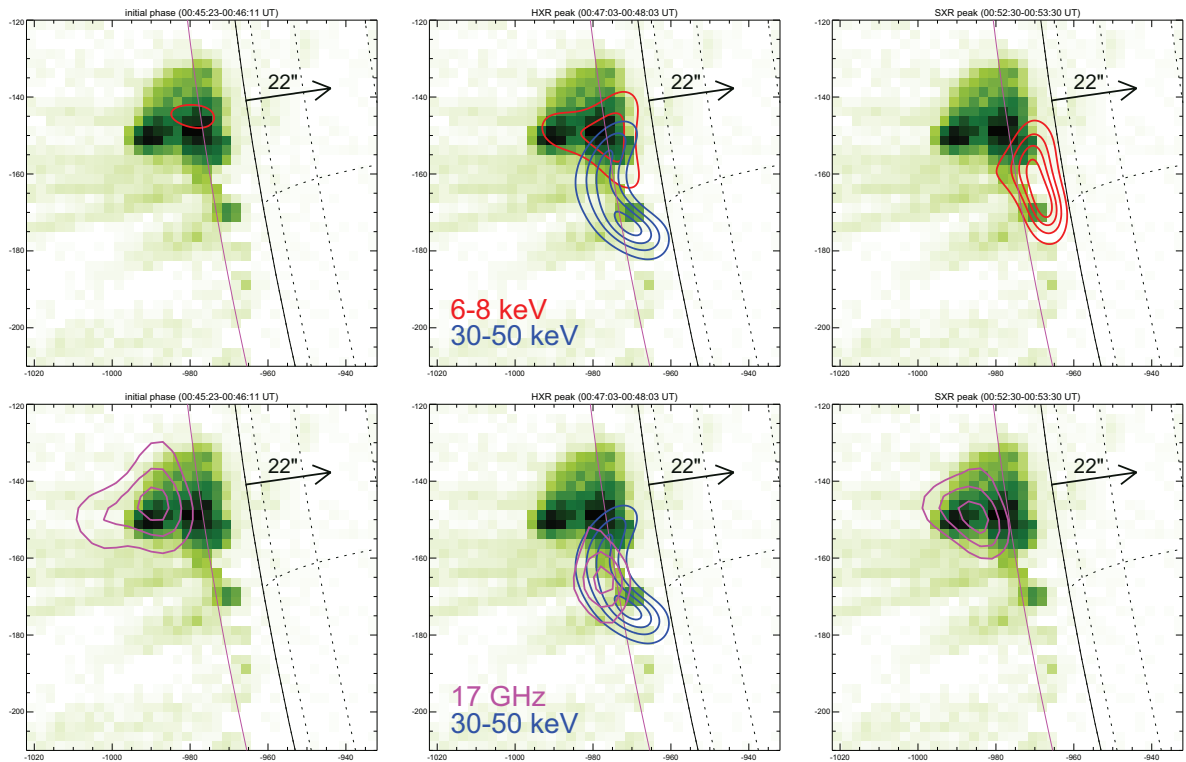}}
\caption{X-ray and microwave imaging during the initial phase of SOL2007-12-31T01:11 (C8.3)  (left), the hard X-ray peak (middle), and the soft X-ray peak (right). Top 
row: Contours in the thermal (6-8 keV, red) and the nonthermal (30-50 
keV, blue) range are shown on an EIT 195 \AA\ image taken
towards the end of the main hard X-ray peak.  The 6-8 keV contour levels are the
same for all time intervals at 20, 40, 60, 80\% of the maximum emission
during the soft X-ray peak time, while the nonthermal emission is shown at 
30, 50, 70, 90\% levels. Bottom row: Microwave contours (magenta) at 17 GHz 
are shown at the 50, 70, 90\% levels on the same
EIT image. For comparison, the 30-50 keV source seen during the
impulsive phase is shown as well (blue). The magenta curve gives the 
location of the limb at 17~GHz, about 12\arcsec\ above the photosphere;
for this reason emission can be seen to lower heights in the HXR images
than in the radio images. The black arrows indicate the occulation
height for the flare location. From \citet{KHG10}.
}
\label{fig:white-071231}
\index{flare (individual)!SOL2007-12-31T01:11 (C8.3)!illustration}
\end{figure}

In this flare (Figure~\ref{fig:white-071231}),
the nonthermal HXR come from a source seen to lie at a height in the
corona that is above the lower-energy thermal soft X-rays.\index{hard X-rays!above-the-looptop source}
This offset was seen previously in the famous ``Masuda'' flare (SOL1992-01-13T17:25, M2.0)
\index{flare (individual)!SOL1992-01-13T17:25 (M2.0)!famous}
\citep{MKH94}\index{Masuda flare}.
NoRH images of the 2007 event show that the 17~GHz emission 
from the flare is co-spatial with the
$>$30~keV source, but the larger occultation
height at radio wavelengths (9~Mm above the optical limb) only allows us
to observe the top part of the nonthermal HXR source at 17~GHz. 
The main flare loop seen during the soft X-ray peak
time is occulted in the 17~GHz observations and emission from the
initial source continues through the soft X-ray peak time. The
microwave spectrum during the hard X-ray peak decreases with frequency as 
expected for gyrosynchroton emission and has a low turnover frequency,
consistent with a coronal origin high above the solar
surface.  Based on the low turnover frequency of 2.5~GHz in the NoRP 
radio spectrum, the magnetic field strength in the 17~GHz source is
estimated to be 30-50~G at a height 25~Mm above the solar photosphere.
\index{magnetic field!height dependence}
Since the electron gyrofrequency for this field strength is only 0.1~GHz, 
emission at 17~GHz is at harmonics in excess of 100 and requires
\index{radio emission!synchrotron}\index{frequency!Larmor}
\index{radio emission!gyrosynchrotron!high harmonics}
relativistic electrons\index{electrons!relativistic}.

In this event the energy power-law spectral index of~3.4 derived from
the radio data is not very different from the value of~3.7 derived from
the \hsi\ spectrum under the thin-target assumptions normally made for
such coronal HXR sources\index{electrons!spectra!HXR/radio comparison}.
Because the occultation height for this event
is so large and the ambient density in the corona is correspondingly low,
extreme number densities of accelerated nonthermal electrons are
required in order to explain the large HXR fluxes. No source of any kind
is visible in the corona at the heights of the flare HXR emission prior
to the event. From the absence of thermal HXR emission at the site of 
the harder photons during the flare, an upper limit of 8~$\times\ 10^9$
\pccm\ can be placed on the ambient density. The observed HXR flux at 50 keV
then requires a density for the nonthermal electrons above 16~keV 
in excess of 10$^9$ \pccm; for a more plausible ambient density of 
$2 \times 10^9$ \pccm at a height of 16~Mm above the photosphere, the
nonthermal electrons would have to outnumber the ambient ions in order
to explain the observed HXR fluxes.
\index{electrons!dominant tail population}

Thus in this event the standard electron-beam scenario where suprathermal electrons move through an
ambient plasma producing hard X-ray emission by bremsstrahlung does not 
work.\index{electron beams!inadequacy of model}\index{flare models!inadequacy of beams}\index{suprathermal populations}\index{beams}
The main problems are that the number of accelerated electrons is 
comparable to the total number of electrons in the pre-flare source and 
that collisional heating would increase the ambient plasma temperature
to superhot temperatures within seconds, producing a bright thermal HXR
source. 
\index{superhot component}
Since no such source is seen, this picture cannot be correct.
Rather, the nonthermal HXR source seems to be produced by a mechanism
that accelerates all electrons and produces an entirely nonthermal electron
distribution.\index{electrons!distribution function!nonthermal}
This suggests that the above-the-looptop source is the acceleration region 
itself\index{above-the-looptop sources!identification with acceleration region}.\index{acceleration region!above-the-loop-top source}
In a purely nonthermal electron distribution\index{electrons!distribution function!purely nonthermal}, collisional losses to electrons are much reduced, making it an efficient hard X-ray source. 
Once a significant fraction of the electrons is accelerated, collisional losses of an accelerated electron are reduced due to a smaller number of ambient electrons. 
This simple picture suggests that once a significant fraction of electrons is accelerated, it might not take that much more to accelerate all of the electrons.
Observations of a similar event in which the pre-flare density\index{density!pre-flare} in the
region of the nonthermal HXR source can be measured 
are needed to confirm this picture.

\citet{ANO07} also report joint \hsi\ and NoRH observations of an
over-the limb M4 flare (SOL2005-07-27T05:02)\index{flare (individual)!SOL2005-07-27T05:02 (M3.7)!radio observations} whose footpoints were
occulted in HXR.
\index{occulted sources}
In that event, a filament eruption preceded the flare and
was clearly visible in the NoRH 34~GHz images; again, nonthermal 
coronal sources\index{coronal sources} are seen in both 25-40~keV HXR and in the microwave
data, but in this event the energy spectra inferred from the two
wavelength ranges differ greatly.\index{filaments!eruptive}
The radio spectrum in this event has a
spectral peak near 10 GHz, i.e., much higher than in the \citet{KHG10}
event, suggesting that effectively none of the radio emission is obscured
from us.

In the \citet{KHG10} event, the radio-deduced and HXR-deduced electron energy
spectra are compatible with one another.
\index{electrons!spectra!HXR/radio comparison}\index{thick-target model!HXR/radio comparison}
It is evident that the discrepancy between the energy spectral indices
derived from the microwave and HXR data in events such as SOL2002-07-23T00:35 (X4.8)
and SOL2003-06-17T22:55 (M6.8) would be largely removed if the
appropriate comparison were with thin-target bremsstrahlung rather than
thick-target: as noted earlier, the actual energy spectrum of electrons in a
thick-target HXR source is flatter than the injected spectrum due to
the faster energy loss rate of lower-energy electrons. 
For the thick-target spectrum 
to be the appropriate comparison, the radio emission from the
flare would have to be dominated by the electrons in the HXR source. 
In active regions where flares occur, 
the filling factor\index{filling factor!surface magnetic field} of
strong magnetic fields at the surface is high and there is little if any
difference between the field strength in the upper chromosphere and that
in the lower corona. 
Therefore, since the volume of the footpoint thick-target HXR sources is so much
smaller than the volume of the coronal loop that connects the footpoints
(the footpoint sources have a depth typically less than 1\arcsec\ at the
top of the chromosphere, whereas a coronal loop can be 20\arcsec\ long), and the incident
nonthermal density at the footpoints will be smaller than the average
nonthermal density in the loop if there is any magnetic mirroring, 
the only way that the electrons in the thick target can dominate the
radio emission is if there are no nonthermal electrons in the coronal
loop, i.e., the nonthermal electrons are entirely confined to the 
chromosphere.
Images of disk flares such as Figure~\ref{fig:white-030617} may seem to
suggest that the locations of the radio emission are consistent with the
HXR footpoints, but such images can be misleading due to their limited
spatial resolution: loops smaller than the resolution of the radio
images cannot be identified as loops in such images, and the distinction
between footpoint sources and loop emission cannot be made. 
In general (see Section~5), radio images of flares are consistent with the paradigm
\index{paradigms!radio dominance of coronal sources}
that the radio emission is dominated by coronal sources.
In any model in which the electrons are accelerated in the corona, there must
always be many more electrons in the coronal volume than in the small
volume occupied by the bremsstrahlung-emitting electrons in the narrow
layer of the chromosphere at the footpoints of the coronal 
loop\index{filling factor!radio/HXR comparison}. 
In this case it remains appropriate to use the the thick-target determination 
of electron energy spectra described in Section~2 for comparison with
the spectral index determined from gyrosynchrotron observations.

\subsection{Oscillations}
\index{quasi-periodic pulsations}

Any periodic or quasi-periodic behavior observed in flares is valuable
because it can reveal the characteristic frequency of a physical process
or environmental property and thus has diagnostic value\index{radio emission!coronal seismology}. 
There have been a number
of events in which such periodicity has been seen simultaneously in
microwaves and hard X-rays, confirming a common origin for the
electrons radiating in the two wavelength domains and providing strong
constraints on the physical processes involved \citep[e.g.,
][]{NKK83,ASI01,GWK03}. \citet{FBG08} discuss another event in which
oscillations are seen both in microwave data from the Owens Valley Radio 
\index{observatories!Owens Valley Solar Array (OVSA)}
Observatory and the Nobeyama Radioheliograph, as well as in the \textit{RHESSI}
hard X-ray light curves. 
\index{observatories!Nobeyama Radio Heliograph (NoRH)}
Periodicities in the range 15-20~s are found
in both wavelength ranges, with degrees of modulation typically around
10\% or lower;
at radio wavelengths, the degree of modulation rises with
frequency, suggesting that it is larger in the optically-thin regime
and thus that it is the number density of nonthermal electrons that is
being modulated.
They also find that the modulations at lower radio frequencies lead the
higher frequencies in time by a small amount, and that modulation of the
circular polarization is in the opposite sense to the modulation of the
intensity: circular polarization is weakest when the intensity is
strongest.\index{polarization!circular}

Another event exhibiting modulations simultaneous in HXRs and radio is
SOL2003-11-03T09:55 (X3.9) \citep{DVL05}. 
\index{flare (individual)!SOL2003-11-03T09:55 (X3.9)!quasi-periodic pulsations}
\index{quasi-periodic pulsations!HXR and radio}
In this event, the
later phase of HXR emission shows a quasi-periodic train of peaks with
a period of order 35~s.
These peaks are clear both in the 150-300 keV HXR
light curve and in the microwave light curves. 
Surprisingly, the decimetric emission at 432 and 327~MHz shows the same modulations\index{quasi-periodic pulsations!decimetric}. 
Images from the \NRH\ at 432~MHz during the modulations show three different sources: 
one close to the limb, just above the HXR source locations in \hsi\ images, 
and two other sources several hundred arcsec higher in the corona.
The higher 432~MHz sources show more pronounced modulations than the lower
source, despite being further from the modulated HXR footpoint sources.
The correlation of the decimetric emission with microwaves and HXRs suggests 
that the emission mechanism is still gyrosynchrotron emission at
decimetric wavelengths\index{gyrosynchrotron emission!Razin suppression}.
Since the Razin effect\index{Razin effect}  is usually thought to
suppress gyrosynchrotron emission at such low frequencies, this
observation suggests that Razin suppression is not
effective in this event, at least for the higher sources, and further
that the gyrosynchrotron-emitting nonthermal electrons have access to
large loops at the height of the 432~MHz sources.
(Note that this event was also remarkable for the fact that it exhibited a Type II radio 
burst\index{shocks!type II radio burst}\index{radio emission!type II burst} from a shock apparently driven by soft X-ray loops rising rapidly into the corona; see Dauphin et al. 2006.)\nocite{DVK06}

Possible interpretations of such oscillations have been reviewed by 
\citet{NaV05} and \citet{NaM06}. \citet{FBG08} test two models with
their event:
oscillations driven by MHD loop oscillations, or quasi-periodic
injections of fresh nonthermal electrons into a stable source.
\index{quasi-periodic pulsations!HXR and radio!interpretation}
An important constraint is that the oscillations are observed to be 
essentially in phase across the whole microwave spectrum, whereas a 
number of MHD modes (sausage, torsional)\index{waves!MHD modes} would
predict that the optically thick (lower) frequencies and optically thin
(higher) frequencies should show opposite behavior. Thus, the authors
conclude that quasi-periodic injections of fresh nonthermal electrons
provide a better fit to the observations.

\subsection{Microflares}
\index{microflares!radio observations}

This topic is discussed at greater length by
\citet{HHB10} and here we briefly mention radio aspects of HXR
microflares.
\citet{KSG06} compared the radio and HXR emission in a sample of
microflares, using \textit{RHESSI} and NoRH data\index{observatories!Nobeyama Radio Heliograph (NoRH)}.
Microflares, X-ray bright-point flares, coronal X-ray jets and active region transient
brightenings are all consequences of energy releases in the corona that
are smaller than traditional flares, but are nonetheless clearly
identifiable phenomena.\index{jets!X-ray}
The sample of 30 microflares studied  
by \citet{KSG06} showed relatively soft (i.e., steep) HXR spectra, which
would lead one to expect that they would not possess sufficient mildly
relativistic nonthermal electrons  radiate strongly at microwave
frequencies. Yet most of the microflares exhibit nonthermal microwave
emission that requires electrons with energies exceeding 100 keV. 
This suggests that the radio-emitting electrons must have a harder energy spectrum
than the HXR-emitting electrons: the imaging results show that
typically the radio and HXR emission are not exactly coincident,
supporting the possibility that the two electron populations are
different. On the other hand, \citet{Nin08a} investigates a sample of
apparently radio-quiet microflares and finds that their HXR properties
are normal for microflares, as in the sample of \citet{KSG06}.
\citet{Nin08a} used less sensitive microwave patrol data for the
microwave fluxes, and it is possible that more sensitive NoRH
observations would have detected nonthermal radio emission.

A prior study by \citet{QLG04} investigated somewhat larger events (up to
\textit{GOES} B2 in soft X-ray class) using \textit{RHESSI} and OVSA data\index{observatories!Owens Valley Solar Array (OVSA)}.
The advantage of OVSA data is that complete radio spectral information is available,
but their detection level of around 1 sfu is not capable of addressing
events weaker in the radio domain.
About 40\% of their sample of about 200 events are detected in
microwaves, with emission up to about 10~GHz and spectral peaks in the
5-10 GHz range: thus their microwave spectra behave like standard
nonthermal gyrosynchrotron spectra and strongly resemble those of
normal flares\index{gyrosynchrotron emission!microflares}.\index{microflares!gyrosynchrotron emission}
Where measured, the radio spectra typically suggest
harder electron energy distributions than the HXR spectra exhibit, and 
some microflares with negligible responses in the \textit{GOES}
soft X-ray range nonetheless can have strong nonthermal microwave
emission, emphasizing again the possible differences between the
lower-energy nonthermal electrons to which HXRs below~25~keV are
sensitive, and the higher-energy electrons responsible for the
microwave emission.

\section{Morphological comparison of microwaves and hard X-rays}\label{sec:white-5}
\index{flares!microwave/HXR morphology}

\begin{figure}[t]
\centerline{\includegraphics[width=\textwidth]{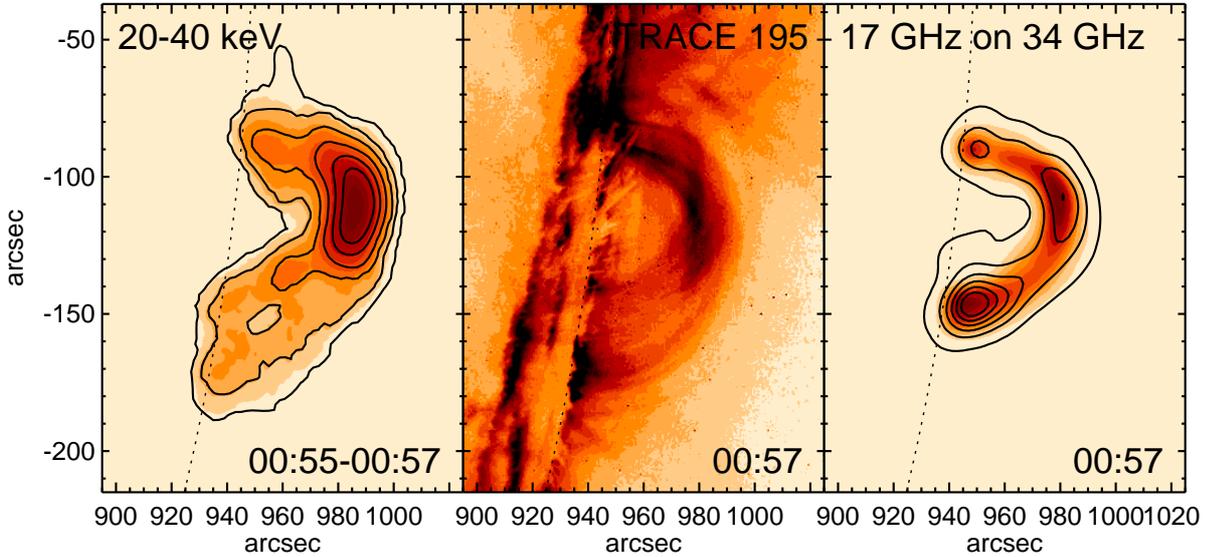}}
\caption{Images of the early phase of the loop flare SOL2002-08-24T01:12 (X3.1) on the west limb. The left panel shows the \textit{RHESSI} image of nonthermal 20-40 keV HXR photons
covering the period 00:55-00:57~UT. 
The middle panel shows a \textit{TRACE}~195~\AA\ image at 00:57~UT, while the
right panel shows contours of the emission at 17~GHz overlaid on a color
image of the 34~GHz emission, also at 00:57~UT. 
Contours in the HXR and radio images are at 5, 20, 35, 50,
65 \& 80\% of the maximum in each image. 
The placement of the
\textit{TRACE} image relative to the other wavelengths is uncertain.
}
\index{flare (individual)!SOL2002-08-24T01:12 (X3.1)!illustration}
\label{fig:white-020824}

\end{figure}
\index{satellites!TRACE@\textit{TRACE}}

\begin{figure}[t]
\centerline{\includegraphics[width=0.9\textwidth]{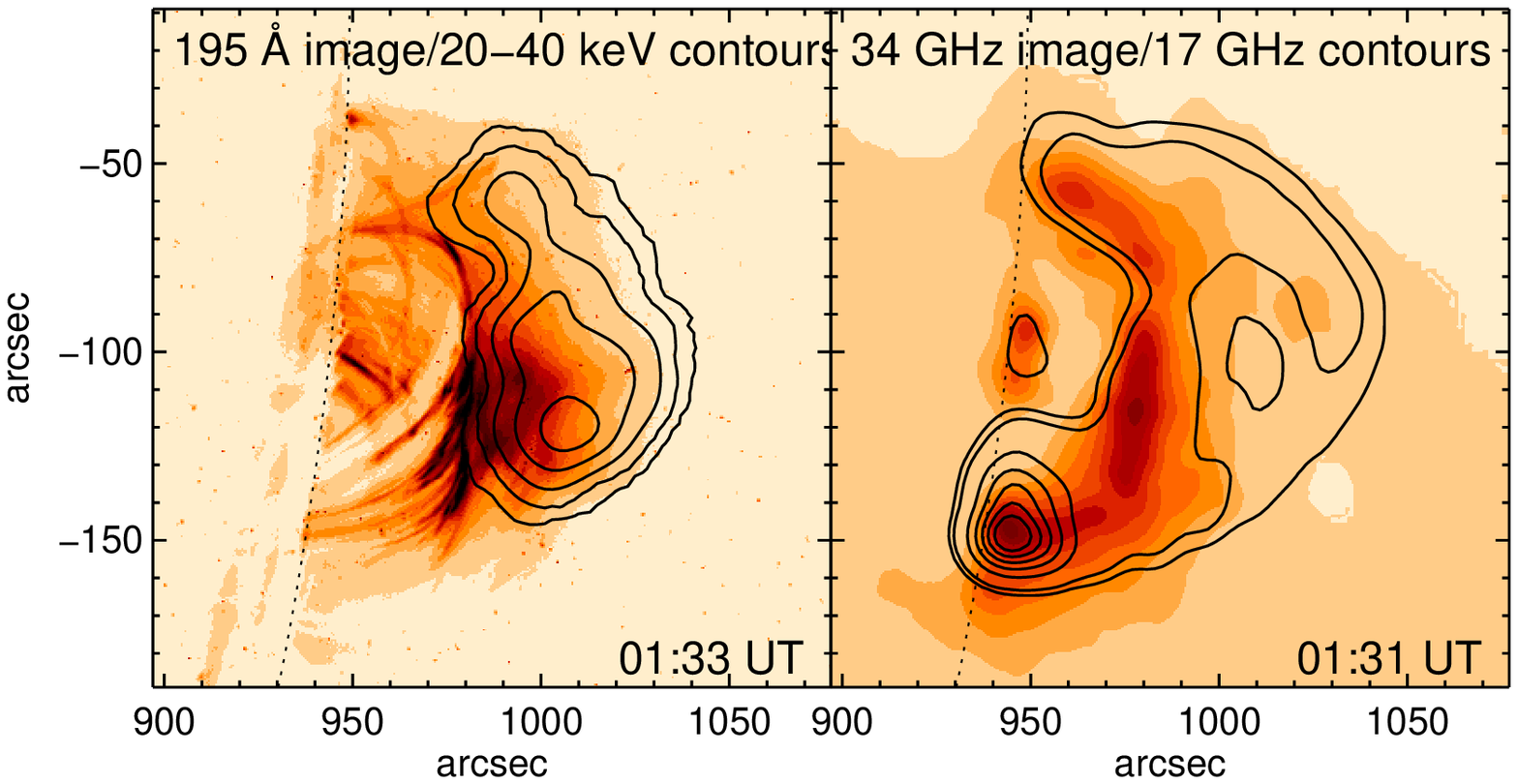}}
\caption{Images of SOL2002-08-24T01:12 later in the event during the
extended gradual phase (near 01:30~UT). 
\index{flare (individual)!SOL2002-08-24T01:12 (X3.1)!radio observations}
At this time the 17~GHz image (contours at 4, 6.5, 20, 35, ..., 80\% of
the maximum, right hand panel) 
shows two loops as well as a bright source at the southern footpoint,
while the 34~GHz image (underlying image in right hand panel) only shows 
the lower loop and a
(relatively) much weaker southern footpoint source. The \textit{RHESSI}
20-40~keV image of nonthermal HXR (contours at 10, 30, 50, 70, 90\% of
maximum, left panel) only shows a high 
coronal source, roughly filling the region between the two loops in the 17 
GHz image, and lying above the cooler (2~MK) loops visible in the 
TRACE~195\AA\ Fe~{\sc xii} image (left panel, image). The 2~MK loops
appear to occupy a smaller volume than the 17~GHz radio source.
}
\label{fig:white-020824-late}
\index{flare (individual)!SOL2002-08-24T01:12 (X3.1)!radio!illustration}
\index{gradual phase!radio!illustration}
\end{figure}

As noted earlier, we have the expectation that radio images will
illuminate the paths of nonthermal electrons along magnetic loops in the 
corona whose footpoints are outlined in HXR by precipitating
nonthermal electrons.\index{image dynamic range!radio vs. HXR}
We have to bear in mind the differences in HXR and
radio imaging data: the best \hsi\ images have a dynamic range of order~100, 
whereas radio images (e.g., from NoRH and the VLA)\index{observatories!Very Large Array (VLA)}
can achieve a dynamic range of order 10$^4$,
and thus the radio images can show secondary sources in locations where
the \hsi\ images do not have enough dynamic range for sources to be
visible above the noise.\index{image dynamic range} 
On the other hand, the brightness of the radio emission is strongly weighted 
by magnetic field strength, and this can lead to different emphases in the
images at the two wavelengths.\index{flares!HXR morphology!weighting by \textbf{B}}

\subsection{``Loop'' flares}
\index{flare types!loop flares}
\index{flares!microwave/HXR morphology!loops}

We expect radio images of flares to outline magnetic field lines carrying
trapped energetic particles: any such particles on open field lines
rapidly move out into the solar wind (it takes 5 seconds for an
electron at speed 0.5$c$ to propagate to 1~R$_\odot$) and are not present in
the regions of strong magnetic fields low in the Sun's atmosphere long
enough to contribute significantly to the microwave emission (but they
may be seen via their plasma emission as Type~III bursts instead).
\index{radio emission!type III burst}
In fact, flare radio images that clearly look like magnetic loops are 
the exception, not the rule\index{radio emission!rarity of loop geometry}. 
This may be partly due to limited spatial resolution: since
NoRH has a typical resolution of order 10\arcsec, any loop needs to be
quite large to appear as such in the radio images. Another factor is
that the magnetic field variation along a loop can emphasize emission
from strong-field regions or regions with $\theta\,\approx\ $90\degr\ 
so much that a loop morphology is not evident.  Accordingly, events 
that do have a loop-like morphology tend to be heavily analyzed.

By far the favorite loop event for radio studies during the \hsi\ period 
is SOL2002-08-24T01:12 (X3.1) on the west limb
\index{flare (individual)!SOL2002-08-24T01:12 (X3.1)!radio observations}
\citep{Kar04,LiG05,Mel06b,TNA08,RZF09,RMS09,NiC09}, which also
produced a fast (1900 \kmps) CME and an energetic particle event at the
Earth.\index{coronal mass ejections (CMEs)!SOL2002-08-24T01:12 (X3.1)}\index{radio emission!loop flare SOL2002-08-24T01:12 (X3.1)}
Figure~\ref{fig:white-020824} shows images in HXRs, microwaves and the
 EUV~195\AA\ band (Fe~{\sc xii} and Fe~{\sc xxiv}) during the early part of the impulsive phase, 
for which \hsi\ data are available. 
The apparent loop has a footpoint separation of order
70\arcsec\ and height of order 50\arcsec. The images provide an interesting
contrast: in HXRs at tens of keV, presumably from nonthermal electrons,
we see a pronounced brightening at the top of the ``loop'' in this early
stage of the flare (here we use
quotation marks to emphasize to the reader that while the source appears
from our line of sight to be a ``single loop''; in fact it could equally
well be an arcade of loops extended significantly east-west, 
but co-located in projection along our line of sight). 
Line-of-sight effects\index{caveats!line-of-sight effects} can also play an important role in the
appearance of the radio source, as discussed in Section~\ref{sec:white-2}.
The flare apparently occurred on the visible disk and
hence footpoint HXR sources should be visible, but they are not
pronounced in these images; there is some evidence for multiple sources
in the region of the southern footpoint at times before 00:55 UT.
The radio images also show a bright peak at the top of
the loop, but at 17~GHz the brightest source is at the southern
footpoint, in a region that appears extended along the limb in the HXR
images. 
The EUV image shows dense cool (Fe~{\sc xii})  and hot (Fe~{\sc xxiv}) material, the
latter also detectable by \textit{RHESSI} via its observations of the Fe~and Fe-Ni X-ray
emission-line complexes \citep{Cthesis}.

\begin{figure}[t]
\centerline{\includegraphics[width=\textwidth]{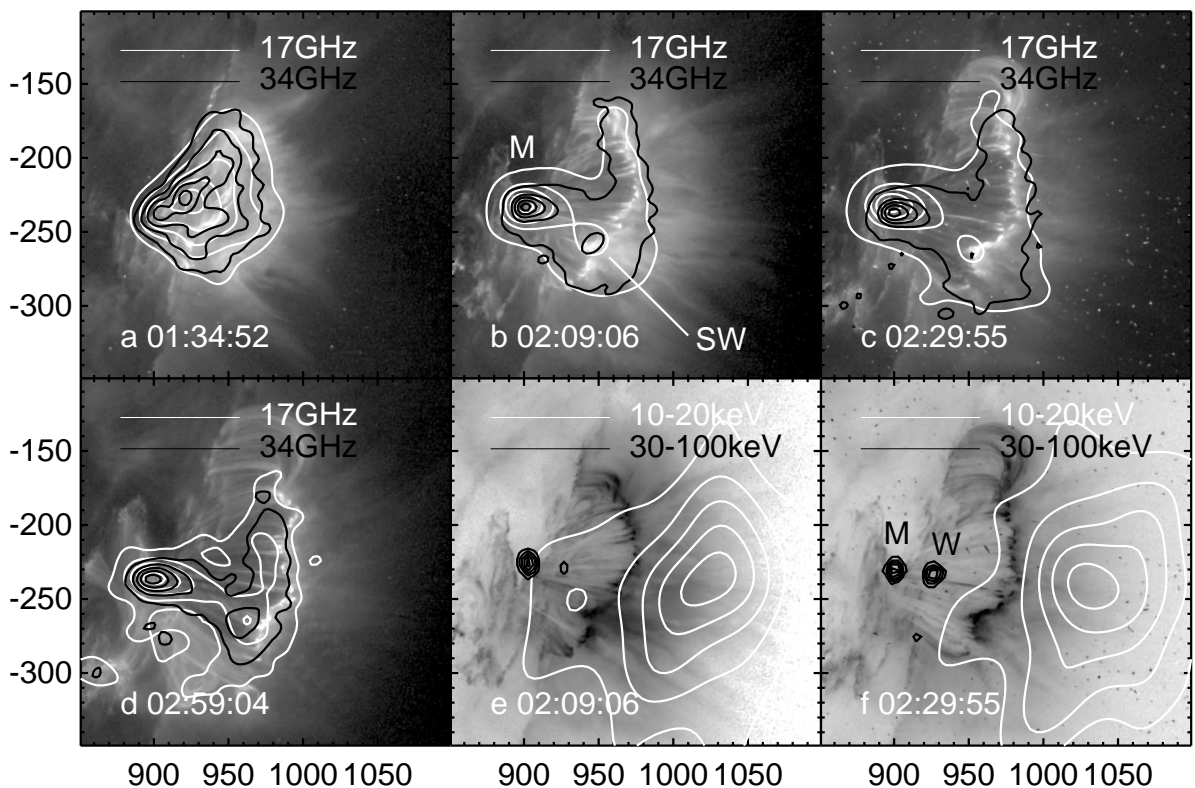}}
\caption{The evolution of the radio and hard X-ray sources later in
SOL2002-08-24T01:12 (X3.1). 
The first four images show a sequence of overlays of
17~GHz (white contours) and 34~GHz (black
contours) images on \textit{TRACE}~195\AA\ Fe{\sc xii}/Fe{\sc xxiv} images during the extended
phase of the flare when the loop system is expanding above the west limb.
The upper four contour levels are at 30, 50, 70 \& 90\% of the maximum in each
image, while the lowest contours are at 10, 4, 2 \& 1\%, respectively, to
match the changes in the peak intensity.
\hsi\ observes the early impulsive phase of the event and then goes into 
eclipse until 02:08~UT, and
the last two images show contours of the \hsi\ 10-20~keV (white) and 
30-100~keV (black) emission overlaid on the \textit{TRACE}~195~\AA\ Fe{\sc xii}/Fe{\sc xxiv} 
images at 02:09:56~UT and 02:29:55~UT (displayed with inverted color table).
The prominent nonthermal radio source is visible at the eastern
base of the radio emission from 02:00~UT onwards. The radio emission from the
loop tops at the northern end of the arcade has a flat spectrum and
hence is
probably thermal free-free emission. The 10-20~keV hard X-rays originate
above the EUV loops, in higher soft X-ray emitting loops, while the
harder 30-100~keV X-rays originate at the same location as the
nonthermal radio source. From \citet{KGW04}.}
\label{fig:white-020421}
\index{flare (individual)!SOL2002-08-24T01:12 (X3.1)!illustration}
\end{figure}

The radio light curve for this event shows a number of intensive peaks\index{quasi-periodic pulsations!SOL2002-08-24T01:12 (X3.1)!}
over a period of 10 minutes starting at 01:00~UT, requiring repeated
injections of energetic particles onto the coronal loops. \citet{RMS09}
analyzed the evolution of the radio morphology and found that each peak
showed the same behavior: during the rise of a peak the southern
footpoint is brightest at 17~GHz, but as each peak decays the loop top
starts to dominate the images. \citet{RMS09} interpret this as requiring
that the highest density of nonthermal electrons always be located at the
loop-top where the magnetic field is weakest, while the nonthermal density 
in the footpoints is initially high but decreases with time. An
implication of this interpretation is that pitch-angle 
scattering\index{pitch-angle scattering!weak at loop top} must
always be weak in the radio-bright region at the loop top.\index{scattering!pitch-angle}
The question arises as to whether these injections occur onto increasingly higher 
field lines, as in the helmet-streamer\index{reconnection}\index{magnetic structures!cusp} reconnection scenario, or onto essentially the same field lines, as expected for a confined flare.
\index{reconnection!helmet-streamer geometry}\index{magnetic structures!helmet streamer}
\citet{LiG05} measured the radius of a circular fit to the radio ``loop'' at 34~GHz and found that the radius of the loop shrank steadily until 01:02 UT, and increased thereafter.\index{magnetic structures!shrinkage}

Later in the flare (01:30~UT; see Figure \ref{fig:white-020824-late}) 
the 17~GHz radio images show two loops
well separated in height, while the 34~GHz images only seem to
show the lower loop. The brightness temperature of the lower loop is
2-3~MK at 17 GHz and around 0.6~MK at 34~GHz, while the upper loop has
a 17~GHz brightness temperature of 1-2~MK but is less than 0.05~MK at
34~GHz. The radio spectrum of the lower-altitude loop is
consistent with thermal emission from a post-flare arcade of loops.
\citet{Kar04} argues that the higher loop, with its
nonthermal radio spectrum, is emitting by gyrosynchrotron emission and 
its appearance requires ongoing acceleration of electrons to nonthermal
energies in the extended phase of the flare. 
The presence of a very
bright source at the southern footpoint, with brightness temperatures
(over 20~MK at 17~GHz, less than 0.6~MK at 34~GHz) indicating nonthermal 
emission, supports this argument, as does the presence of significant
emission in the 20-40~keV range in the same volume as the nonthermal
radio emission (Figure~\ref{fig:white-020824-late}, left panel).

Loop morphologies can also show up in thermal emission. \citet{BBF07}
discuss the development of a bright compact loop which has a flat radio
flux spectrum and thus appears to be
radiating by thermal bremsstrahlung. 
\index{bremsstrahlung!and densities inferred from radio}
Densities of close to
10$^{11}$ cm$^{-3}$ are inferred. Another flare occurs in the same set 
of loops 30 minutes later, and \citet{VeB04}, from analysis of the \textit{RHESSI}
hard X-ray spectra, inferred that the flare must have occurred in a
very dense environment.
The subsequent analysis of the radio data by
\citet{BBF07} confirmed this conclusion. Further discussion of this
topic may be found in the chapter by \citet{FDH10}.

\begin{figure}[t]
\centerline{\includegraphics[width=\textwidth]{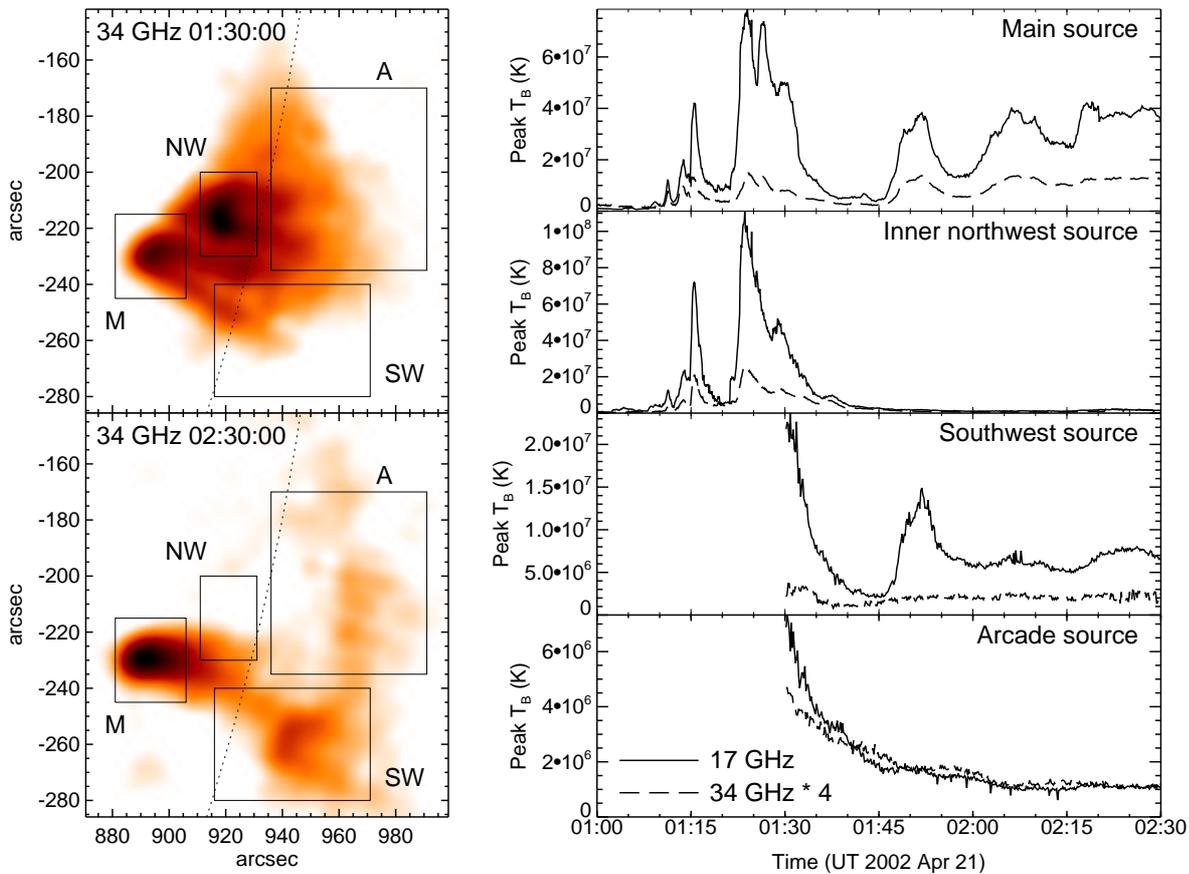}}
\caption{Light curves for individual locations in SOL2002-04-21T01:51 (X1.5). 
The 34 GHz~NoRH images in the left hand panels show the
locations of the four regions chosen for investigation relative to the
morphology of the source at two different times, while the four panels on the
right show the evolution of the maximum brightness temperature in each
of the regions. The light curves show both 17~and 34~GHz curves: the 34
GHz brightness temperatures are multiplied by 4, so that they should be
the same as the 17~GHz values if the emission is optically-thin thermal
emission (as is clearly the case for the arcade source).}
\index{flare (individual)!SOL2002-04-21T01:51 (X1.5)!illustration}
\label{fig:white-020421lc}
\end{figure}

\subsection{ SOL2002-04-21T01:51 (X1.5)}

The well-studied limb flare SOL2002-04-21T01:51 (X1.5) exhibited a spectacular arcade of
post-flare loops in EUV images.
\index{flare (individual)!SOL2002-04-21T01:51 (X1.5)!radio observations}
\index{flare (individual)!SOL2002-08-24T01:12 (X3.1)!radio observations}
\citet{KGW04} studied the radio emission from this flare and found, much as in SOL2002-08-24T01:12 (X3.1), that bright nonthermal sources were seen at a
number of locations low in the corona\index{arcade!non-thermal radiation}, while the top of the post-flare arcade clearly shows thermal bremsstrahlung emission from the dense
plasma there\index{arcade!microwave free-free emission}. 
Distinct regions of radio emission with very different time
behavior can be identified in the radio images, and in
particular a peculiar nonthermal source seen in radio and hard X-rays low in the corona at the
base of the arcade is seen to turn on at 01:45~UT, some 
30~minutes after the start of the impulsive phase (Figure \ref{fig:white-020421}). 

This event is striking for the wide range of
types and locations of both radio and hard X-ray emission it displays.
The energy distribution inferred for the radio-emitting electrons
during the impulsive phase is quite similar to that inferred for the
hard X-ray-emitting electrons, and the radio and hard X-ray light
curves show similar time structure. 
In the radio images we can identify
at least four spatially-distinct regions that show quite different
temporal behaviors (Figure \ref{fig:white-020421lc}). 
The brightest radio emission in the flare comes from a
location to the north-west of the site where the flare starts,
under the middle of the loop arcade, but the radio emission from this
location fades after the impulsive phase and is unimpressive during the
extended decay phase. On the other hand, the main source and the thermal
and nonthermal sources in the arcade (distinguishable by their
brightness temperature spectra) all participate in the extended
phase. 
This indicates that energy release is ongoing throughout this
phase and apparently distributed throughout the coronal volume above the
flare site\index{flares!energy content!distributed coronal release}.

\section{Millimeter- and submillimeter-wavelength emission from
flares}\label{sec:white-6}
\index{radio emission!mm- and submm-wavelengths}
\index{radio emission!THz band}
\index{submillimeter emission!relevance to $\gamma$-rays}
\index{gamma-rays!electron distribution common to submm band}

Radio emission from solar flares at millimeter and shorter wavelengths
is expected to have two main characteristics: since optically-thin
thermal emission has a flat flux spectrum, it should become increasingly
important relative to the falling spectrum of nonthermal gyrosynchrotron
emission as one goes to higher frequencies. However, higher-frequency
nonthermal gyrosynchrotron emission requires emission at ever higher
harmonics of the electron gyrofrequency\index{frequency!Larmor!harmonics}, and this makes it sensitive to
electrons with much higher energies than are required for microwave
emission \citep[MeV and higher energies rather than tens to hundreds of
keV:][and Figure~\ref{fig:white-0}]{Ram69,WhK92,RSE94}. 
Bremsstrahlung\index{bremsstrahlung!efficiency decrease with frequency} 
is increasingly inefficient as
electron energy increases because faster electrons suffer less
deflection by a nucleus. However, synchrotron 
emission\index{gyrosynchrotron emission!efficiency increase with frequency} increases in
efficiency as electron energy increases.
Thus millimeter-wavelength observations are a very
sensitive diagnostic of $\gamma$-ray-emitting electrons\index{gamma-rays!less sensitive than mm waves},
more sensitive, in fact, than current $\gamma$-ray detectors in the sense
that nonthermal emission from smaller flares is more easily detected at 
millimeter wavelengths \citep[e.g., ][]{KWG94,Whi94,SWL96,RWK99,Whi99}.

During the period of \textit{RHESSI} observations, the Solar Submillimeter
Telescope  \citep[SST;][]{KLC08} has routinely monitored the Sun at 212~and 405~GHz 
from the Argentinian time zone\index{observatories!Solar Submillimeter Telescope (SST)}.
Typically only major flares can be
detected at these wavelengths, where the atmosphere has a major
influence on detectability (the sky opacity at 405~GHz is generally in
excess of~1).
\index{absorption!Earth's atmosphere!mm waves}\index{submillimeter emission!atmospheric transmission}

\begin{figure}[t]
\centerline{\includegraphics[width=0.8\textwidth]{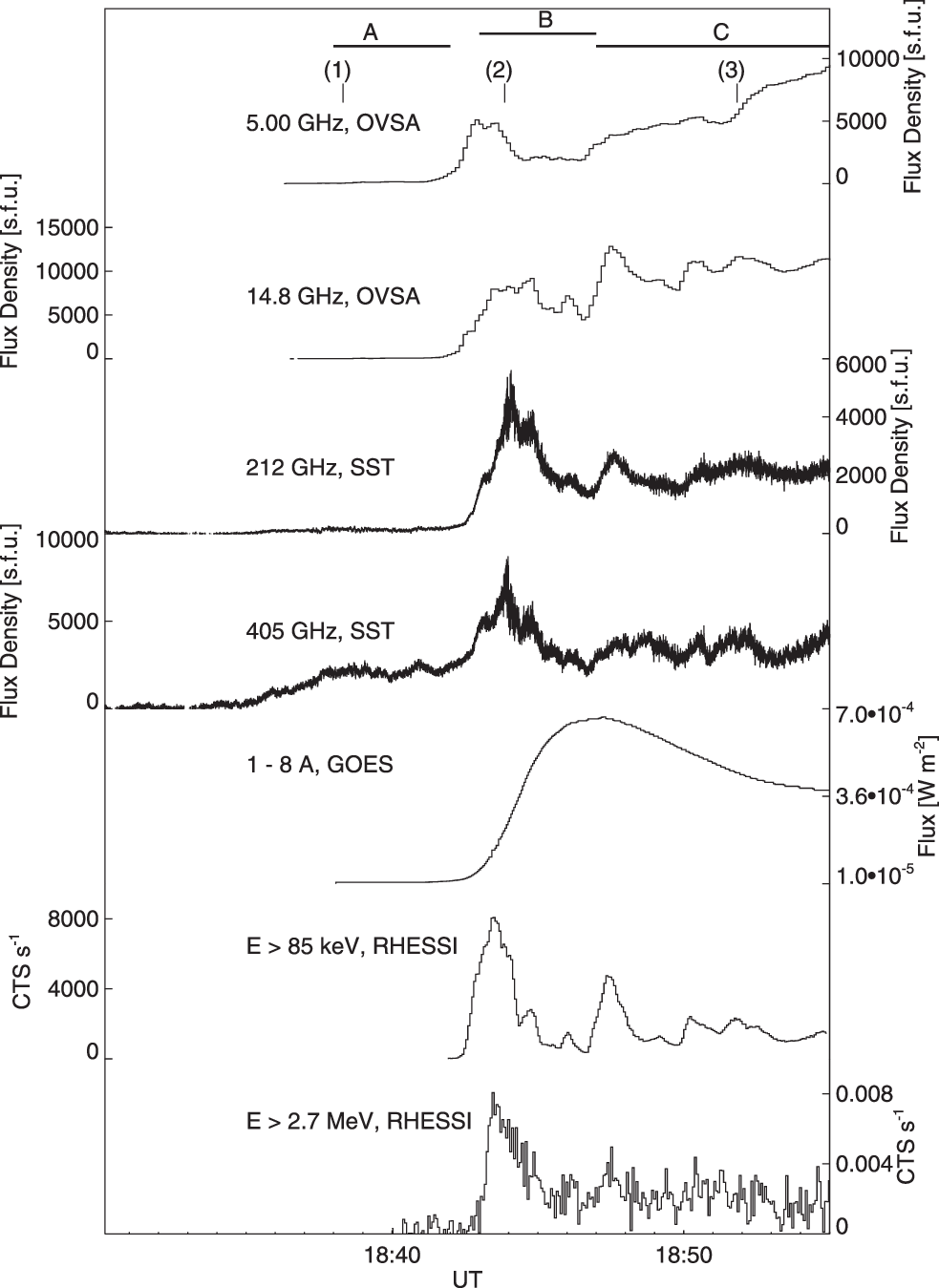}}
\caption{Light curves for microwave, millimeter, submillimeter, hard
X-ray and $\gamma$-rays from SOL2006-12-06T18:47 (X6.5) \citep[from][]{KTG09}.
\label{fig:white-sst_061206}}
\index{flare (individual)!SOL2006-12-06T18:47 (X6.5)!illustration}
\end{figure}

An example of SST data is shown in Figure \ref{fig:white-sst_061206}, 
from SOL2006-12-06T18:47 (X6.5).  
\index{flare (individual)!SOL2006-12-06T18:47 (X6.5)!submillimeter emission}
This event is typical in that there is in general a good correlation between the radio emission and
the HXR/$\gamma$-ray emission, consistent with the picture of the radio
emission resulting from gyrosynchrotron emission by the same nonthermal
electrons that produce the HXR/$\gamma$-ray emission by bremsstrahlung.
However, this event, in common with a number of others, shows a major
inconsistency with this picture: the radio spectrum shows a normal
nonthermal behavior with a spectral peak in the 10-20~GHz range, with a
falling spectrum at high microwave frequencies, but the SST data show an
increasing spectrum from 212~to 405~GHz. 
Such behavior is not understood\index{submillimeter emission!increase at high frequencies}.
 Other events detected by SST show radio
spectra that are consistent with a single nonthermal electron
population, as in the case described by \citet{GTS09}: in that flare 
no flux was detected above 250~keV, yet the 212~GHz emission was
clearly detected.

As discussed earlier, it is well known that there is generally a discrepancy
between the energy distribution of the nonthermal electrons that produce
the HXRs and those that produce the radio emission at high frequencies:
the radio data generally show a harder spectrum \citep[e.g.,
][]{KWG94,RWK99,SWG00}\index{electrons!spectra!HXR/radio comparison}\index{spectrum!upwards breaks}.
This is attributed to the fact that the energies of electrons emitting 
the HXRs (many tens of keV) are smaller
than those of the radio-emitting electrons (hundreds of keV): if the
typical electron energy spectrum breaks up at energies above a few hundred keV,
this could explain the observations. 
In principle, $\gamma$-ray spectra 
should show a break up at higher photon energies if the electron energy
spectrum has such a break, but in the relevant energy range (photon
energies above 500~keV) it is 
difficult to separate the electron bremsstrahlung spectrum from the
nuclear line spectrum that usually dominates $\gamma$-rays above 1 MeV. 
There is one class of events for which the the electron bremsstrahlung
spectrum above 0.5~MeV is more easily determined, the so-called
``electron-dominated events'' \citep{MRK91}\index{flare types!``electron-dominated''}. 
These events result from
extremely hard electron energy spectra whose bremsstrahlung dominates the
usual nuclear line region. The prototype of this class of event was
SOL1989-03-06T13:56 (X15)\index{flare (individual)!SOL1989-03-06T13:56 (X15)!gamma-ray spectrum}, in which the photon spectrum during the main emission peak had a
spectral index of 2.5 from 0.3-0.8~MeV, flattening to about 1.4 from 1~MeV 
to 10~MeV \citep{MRK94}. \citet{PMM94} argued that the entire 0.3-10~MeV 
photon spectrum in this event can be explained by bremsstrahlung from a
single electron energy 
distribution with power-law index~2.2. 
In their model, the steeper photon
spectrum below 1~MeV is due to the fact that the lower-energy photons are
emitted nearly isotropically whereas the higher-energy photons are
preferentially beamed along the direction of electron motion 
into the Sun \citep[e.g.,][]{VFC87}. 
As noted in Section~2.2,
relativistic and electron-electron effects in a single power-law energy 
distribution can produce a flattening of order 0.5 in the photon 
spectral index above 500~keV. \citet{TVB98} discuss the
electron-dominated flare\index{flare types!``electron-dominated''} 
SOL1990-06-11T09:43 (M4.5), which exhibited a flattening of
\index{flare (individual)!SOL1990-06-11T09:43 (M4.5)!gamma-ray spectrum}
the bremsstrahlung photon spectrum at high energies but also major variations
in the spectrum with time. During the brightest $\gamma$-ray peak the spectrum
flattened by about 1 at 700~keV, whereas during other emission peaks the
lower-energy photons had a spectrum steeper by 2-2.5 than the spectrum above
400 keV. The radio spectrum from 35 to 50~GHz during the brightest peak
\index{flare (individual)!SOL1990-06-11T09:43 (M4.5)!radio emission}
corresponds to an electron energy distribution with index of order~3,
as does the bremsstrahlung photon spectrum above the break, so in this
event they appear to be compatible with a single electron energy
distribution. \citet{VTB99} study another electron-dominated\index{flare types!``electron-dominated''} event in which the bremsstrahlung photon spectrum flattens above 500~keV by about~1 in the spectral index. 

But even a break upwards in the electron energy spectrum, as described
above, cannot produce a radio spectrum that rises in the submillimeter range\index{submillimeter emission!increase at high frequencies}.
One can envisage a source of ultra-relativistic leptons\index{particles!ultra-relativistic leptons} 
in a very high magnetic field region: \citet{TKL08} suggest that
ultra-relativistic positrons resulting from pion decay are a possible source of
the impulsive submillimeter component in SOL2003-10-28T11:10 (X17.2)
\index{flare (individual)!SOL2003-10-28T11:10 (X17.2)!submillimeter emission} 
seen in conjunction with high-energy protons ($>$200~MeV), although it seems
unlikely that sufficient positrons can be produced by this means and
energetic electrons are still a more plausible interpretation.
An optically-thick thermal source could produce a rising spectrum, but it 
would have to be either exceptionally large or exceptionally
hot \citep[e.g., ][who prefer an interpretation in terms of gyrosynchrotron 
emission from electrons in a very strong magnetic field]{SSM07}. 
\index{flare (individual)!SOL2006-12-06T18:47 (X6.5)!submillimeter emission}
\index{observatories!Solar Submillimeter Telescope (SST)}
In the case of SOL2006-12-06T18:47 (X6.5)\index{flare (individual)!SOL2006-12-06T18:47 (X6.5)!submillimeter emission}, the SST observations place an
upper limit of 15\arcsec\ on the submillimeter source size\index{submillimeter emission!angular resolution limit}, and for a
source that small to produce the observed flux via thermal emission 
requires such high
temperatures that the source would be intense in the \textit{GOES} soft X-ray range, 
and would have been seen at other wavelengths \citep[e.g., ][]{KTG09}. 
At the time of writing, there is no completely accepted explanation for this 
spectral feature; a more detailed discussion of this topic may be found
in \citet{KBC10}.

\section{Decimeter and low-frequency radio emission in association with
hard X-rays}\label{sec:white-7}
\index{radio emission!decimeter}

Most decimeter (300-3000~MHz) and 
low-frequency (here referring to frequencies below 300~MHz) emission 
from the Sun is dominated in flares by very bright plasma emission\index{radio emission!coherent!plasma emission},
i.e., conversion of electrostatic Langmuir waves at the electron plasma
frequency ($f_p\ =\ 9000 n_e^{1/2}$) into electromagnetic radiation at
the fundamental $f_p$ and the second harmonic 2$f_p$.\index{frequency!plasma}
\index{plasma instabilities!bump-on-tail}
The Langmuir waves\index{waves!Langmuir}
are generated by coherent processes\index{radio emission!coherent!Langmuir waves}, such as a bump-on-tail
instability\index{plasma instabilities!bump-on-tail} driven by an electron beam, and so this mechanism can reach 
very high brightness temperatures even though relatively few electrons are
involved. 
\index{radio emission!high brightness temperature}
Furthermore, by definition, low-frequency plasma emission
comes from regions of low density, whereas hard X-ray emission by
bremsstrahlung from collisions is preferentially seen from regions of
high density. 
\index{hard X-rays!and decimeter emission}
For this reason, 
we might not expect a close relationship between low-frequency radio
bursts and hard X-ray emission: the electrons producing both have
similar energies (plasma emission can be very strong from electrons with
energies of tens of keV), but they may be located in very different
environments. As an example, \citet{VKT03} discuss the radio and HXR
data for an M8 flare in which there are multiple HXR sources and
multiple spatially-distinct radio sources in the 100-500~MHz range, but
little spatial correlation between the two wavelengths: in particular
the early HXR emission in the event seems to come from compact closed
magnetic loops with little escape of energetic electrons onto the higher
field lines where we believe that the low-frequency radio sources are
located. In a later phase of this event a new HXR source low in the atmosphere 
is associated with the appearance of new radio sources at much greater
heights, suggesting simultaneous injection of electrons over a wide
range of spatial scales.

\subsection{Radio bursts and energy release in flares}

\citet{BGC05} and \citet{ArB05} have carried out surveys of the types of
radio bursts in the range 0.1-4~GHz that occur in flares that exhibit 
hard X-rays, using data from the {\it Phoenix-2} spectrometer operated by
ETH-Z{\"u}rich. 
\index{radio emission!coherent}
\index{observatories!Phoenix-2}
This frequency range is dominated by plasma emission and 
other forms of coherent processes such as electron cyclotron maser emission\index{radio emission!coherent!electron cyclotron maser}.

\citet{BGC05} investigated 201 flares and found that about 20\% of such
flares did not exhibit radio emission in their range \citep[although
many of the events with no radio emission were close to the limb and
that fact may play a role in their non-detection; see][]{BBM07}, but over half of
the HXR-productive flares followed a similar pattern: at lower
frequencies Type~III bursts (indicating electron beams) are seen
propagating outwards, and at the highest frequencies the bottom of the
optically-thick gyrosynchrotron continuum from nonthermal electrons is
seen, while in between (typically in the range 500~MHz to 2~GHz) pulsations
and/or narrowband spikes (often attributed to electron
cyclotron maser emission) may occur.\index{quasi-periodic pulsations}\index{radio emission!decimetric spikes}\index{beams!and type III bursts}
Note that a plasma frequency of 1~GHz
corresponds to an electron density of 1.2 $\times\ 10^{10}$ \pccm\
and an electron cyclotron frequency of 1~GHz corresponds to
a magnetic field strength of 360~G, both of which
are values characteristic of the low corona and heights where we might expect
energy release to be taking place\index{radio emission!coherent},
For this reason, there has long been
interest in understanding the relationship between energy releases and 
coherent radio bursts in the decimetric range \citep[e.g.,
][]{AWB85,ABK90,GAB91}. 

\citet{ArB05} looked in more detail at the relationship between the
timing of radio and hard X-ray emission in this frequency range. They
found that only about 20\% of type~III bursts show a close temporal
correlation with HXR peaks, generally trailing the HXR by about 0.5
seconds\index{radio emission!type III burst}\index{radio emission!type III burst!reverse slope}.
Some reverse-drift and patchy bursts also show such delays
relative to the HXRs. 
Such delays can be explained by scattering of the radio
emission in the inhomogeneous medium around the source.\index{scattering!radio emission}

Narrowband decimetric spikes show the strongest association with HXRs,
in the sense that 95\% of such spikes occur during HXR emission
\citep[but only 2\% of HXR flares show spikes;][]{GBA91}\index{radio emission!narrowband decimetric spikes}.
The spikes have timescales shorter ($<$0.1~s) 
than we can easily measure in hard X-rays, so timing comparisons are
difficult. In addition, spikes may be occurring across a range of frequencies
at any given instant, so choices have to be made when making comparisons
with HXR light curves. \citet{DaB09} carry out a careful comparison of
decimetric spikes and HXRs in the time domain.  They cross-correlated the 
radio spike and \textit{RHESSI} HXR light curves to measure delays. 
The mean correlation coefficient was around 0.7, which is statistically 
significant for the size of the sample.  The delays showed
a broad distribution with a standard deviation of order 4~seconds
but a mean consistent with zero delay.  

\citet{BaB09} investigated the relationship between spike and HXR
sources by comparing the spatial locations of the two emissions for a flare
at the limb\index{radio emission!decimetric spikes!and hard X-rays}\index{hard X-rays!and decimetric spikes}.
The spike locations were obtained from \NRH\ observations between 300
and 432 MHz, while \textit{RHESSI} provided HXR images. The radio spikes were found to
be significantly displaced from the HXR sources and at a greater height.
Low-frequency radio emission is typically scattered and refracted in the corona,
but any refraction will make the apparent height lower than the true
height \citep[e.g.,][]{MeD88}. It is possible that there is HXR emission from the spike
locations but that it is much weaker than the lower-altitude sources
and is lost in the limited dynamic range of the HXR images. 
An occulted event would be needed to test this possibility.\index{occulted sources}
\citet{NCH09} investigate some features in decimetric dynamic spectra that
they argue, based on simultaneous \textit{RHESSI} data, represent 
upflows due to heating of the chromosphere by nonthermal electrons.\index{upflows!and radio dynamic spectra}

\begin{figure}[t]
\centerline{\includegraphics[width=0.95\textwidth]{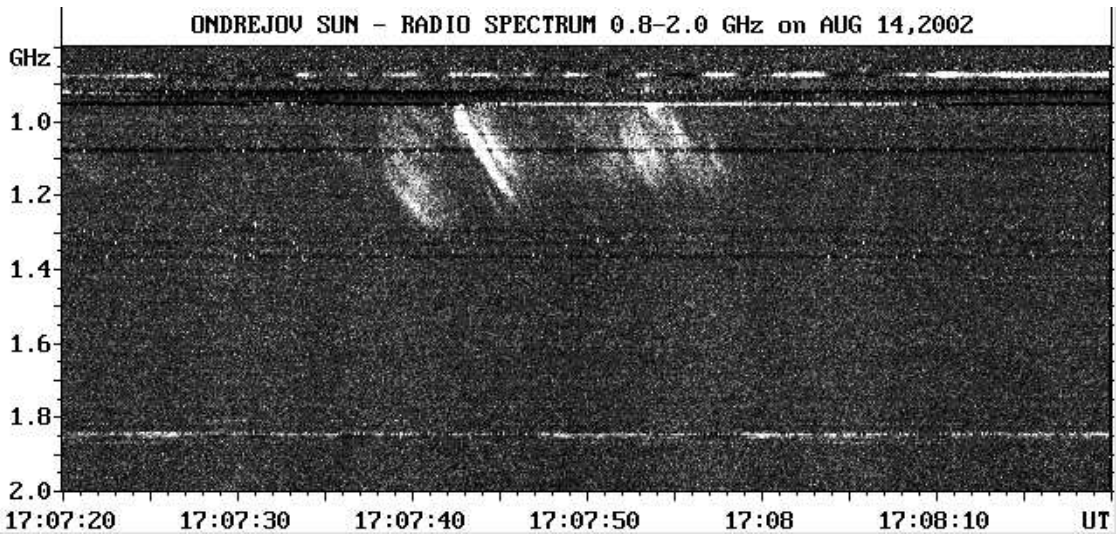}}
\caption{A dynamic spectrum of reverse-drift radio bursts observed
with the Ondrejov spectrograph between 1.0~and 1.2~GHz \citep{FaK07}.}
\label{fig:white-revdrift}
\index{radio emission!reverse drift!illustration}
\end{figure}

\subsection{Hard X-rays from reverse-drift bursts}
\index{radio emission!type III burst!reverse drift with HXRs}

Electron beams are a common phenomenon in the solar corona (even in the
absence of flares), visible
through both their plasma emission at radio wavelengths and their 
bremsstrahlung hard X-ray emission if they reach the chromosphere\index{beams}\index{reconnection!and electron beams}
The exact mechanism by which these electron beams are generated is still
unknown. 
For example, X-point reconnection is believed primarily to produce
oppositely directed bulk ion flows, but not collisionless electron
beams, and so the production of such beams in this mechanism would
require a secondary process. 
Detection of both radio and HXR emission from the same feature
would be an important step in understanding their origin, since the HXR
data reveal the energy distribution of accelerated electrons and the
radio data have the potential of revealing the location of the
acceleration site and the physical conditions therein.

While the most common manifestation of electron beams is the 
Type~III radio burst\index{radio emission!type III burst} (discussed next), generated by electrons moving
outwards through the corona on open field lines, symmetry suggests that
there should be roughly equal numbers of electron beams going downwards.\index{radio emission!type III burst}\index{radio emission!type III burst!reverse drift}\index{beams!and type III bursts}
If such beams generate plasma emission, they will be seen to
drift from lower to higher frequencies as time proceeds, the reverse of 
Type~III bursts, and hence they are referred to as ``reverse-drift'' bursts.
Since the electrons in such events are moving downwards towards the
chromosphere, they seem more likely to be associated with hard X-rays.
At frequencies above 1~GHz, such bursts, if due to plasma emission, must
arise low in the corona and propagation times down to the chromosphere
are typically less than 0.5~seconds.\footnote{Historical note: an early
paper on flare HXR bursts by \citet{AnW62} noted the similar timing of 
Type~III bursts
and suggested that they may be due to the same electrons that produced
the hard X-rays. The more
likely connection to downwards-moving reverse-drift bursts, and the
absence of such bursts in that event, was pointed out by \citet{Kun63a}, 
with further discussion presented in \citet{Kun65}.}

The relationship between reverse-drift bursts and hard X-rays has been
studied by \citet{KFK04} and \citet{FaK07}. An example is shown in
Figure \ref{fig:white-revdrift}. 
From comparison with hard X-ray data\index{hard X-rays!and reverse-drift type III bursts}
in over 20 groups of reverse-drift bursts,
these studies find that bursts are mostly
observed during the rise phase of the hard X-ray emission, but in the
range above 1~GHz there was no one-to-one
relationship between individual HXR peaks and individual reverse-drift
bursts at a time scale of order 1~second. By contrast, 
in the frequency range below 1.4~GHz \citet{ABD95}  found 
correspondence between individual X-ray peaks and
fast-drift radio bursts (both Type IIIs and reverse-drift bursts) 
at a level of 26\%\index{radio emission!type III/HXR correlation}. 
Thus the higher-frequency (and presumably,
higher-density environment) reverse-drift bursts above 1~GHz 
currently do not support the idea that such downgoing electron beams
visible at radio wavelengths are
responsible for the bulk of the HXRs emitted by flares.

\subsection{Hard X-rays from Type III radio bursts}
\index{hard X-rays!and type III bursts}
\index{radio emission!type III burst!and hard X-rays}

Depending on the height at which they originate, Type III radio bursts may 
seem to be
unlikely candidates for correlation with hard X-rays. Those originating 
at densities below 10$^9$~\pccm\ ($f_p$ = 300 MHz) do not experience
much column density as they travel away from the solar
surface, making it difficult for them to produce detectable
bremsstrahlung HXRs as they propagate outwards. 
But if they have downgoing 
counterparts, not necessarily detected as radio sources, then the downgoing electrons
will encounter a greater column density and thus could produce
observable HXRs.\index{beams!downgoing!radio detectability}

Type~III bursts have a high degree of correlation
with the onset of HXR emission in impulsive flares\index{impulsive phase!radio type III bursts}: in about 
30\% of flares (and a larger proportion of impulsive flares) Type~III
emission occurs as the impulsive phase begins \citep[e.g.,
][]{CMR86,CaR88a,ABK90}, but typically these impulsive-phase radio bursts 
do not last for the entire period of
HXR emission so a one-to-one association with HXR
features cannot be established\index{hard X-rays!radio emission!type III/HXR correlation}.

\citet{SKC09} carry out a detailed study of the quantitative conditions
needed for outward-traveling electron beams such as
Type-III-emitting streams to be detectable as thin-target nonthermal HXR
sources. 
\index{bremsstrahlung!thin-target}
\index{models!coronal electron density}
They find that radially-extended HXR sources can indeed be
produced by electron beams propagating through standard coronal density
models, but only if the electron beams are intense: they find that 
10$^{35}$ electrons above 10 keV are needed for
detection of the source, but that 10$^{36}$ electrons above 10 keV are needed
if \textit{RHESSI} is to be capable of imaging the HXR source. Such strong beams
are more characteristic of the downward flux at an HXR-emitting footpoint
in a flare than of the outward-propagating Type-III-burst electron beams 
as inferred from measurements in the solar wind \citep{KKC07}. These
calculations assume that the electron beam is generated in a region of
density $3\,\times\,10^9$ cm$^{-3}$ (plasma frequency of 500~MHz) in a
density model with scale height $\sim$10$^{10}$ cm; beams
generated at lower starting densities will experience less column
density and therefore will be difficult to detect unless the beams are
much stronger.

\begin{figure}[t]
\centerline{\includegraphics[width=0.95\textwidth]{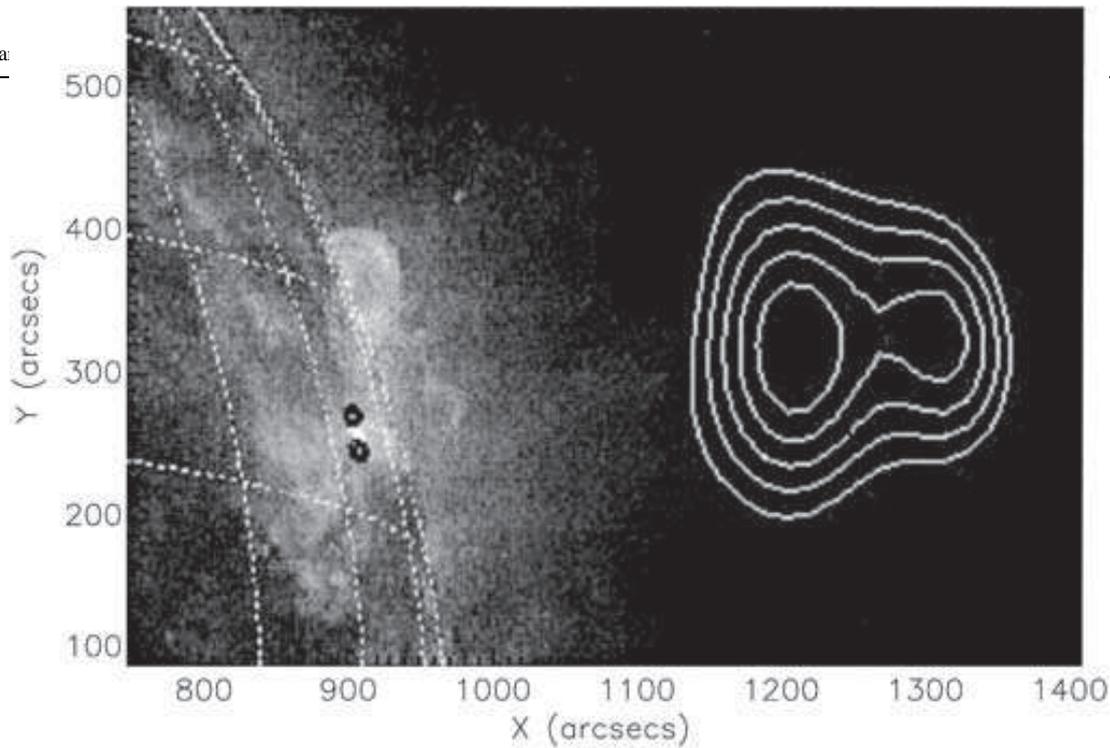}}
\caption{\textit{RHESSI} iso-contours (black) (40, 60, 80\% of the maximum) at
25-40 keV and NRH contours at 410 MHz (white) (50, 60, 70, 80, 90\%) 
for the SOL2002-02-20T11:07 (C7.5) \citep{VKL02}. 
The \textit{RHESSI} and NRH contours are superposed on
an EIT image obtained at 11:12 UT.
}
\label{fig:nrh_typeiii}
\index{radio emission!type III burst!illustration}
\index{flare (individual)!SOL2002-02-20T11:07 (C7.5)!illustration}
\end{figure}

\citet{VKL02} compared Type III bursts
with hard X-rays in one of the first flares observed by \textit{RHESSI},
SOL2002-02-20T11:07 (C7.5).
\index{flare (individual)!SOL2002-02-20T11:07 (C7.5)!type III bursts}
In that event the radio emission from outgoing
electron beams seems to be very well correlated with hard X-ray
emission.\index{beams!radio/HXR correlation}
Figure \ref{fig:nrh_typeiii} shows the relative locations of
the HXR sources and 410~MHz emission observed with the
 \NRH\index{observatories!Nan{\c c}ay Radioheliograph}. 
The \textit{RHESSI} images show three separate sources in
locations consistent with thick-target footpoint emission, each with a
different temporal behavior (only two are visible in the figure)\index{radio emission!type III burst!and hard X-rays}.
At one point there is a transition in which
the source initially brightest in HXRs fades and is overtaken by a
different source.
This transition occurs at a time when the 410~MHz
radio emission shows a distinct brightening, suggesting a causal
connection between the radio and HXR emission. 
\NRH\ radio images of this event
also show a transition.
Initially the HXR footpoint sources occupy a
spatial scale of order 20\arcsec\ while the 410~MHz radio emission shows
two sources aligned almost radially above the flare site at altitudes of
order 300\arcsec\ \& 400\arcsec\ above the photosphere, with the lower
source being the brighter: after the transition in the location of the
brightest HXR emission, the higher 410~MHz source becomes brighter.
\citet{VKL02} interpret the good correlation between the HXR and radio
emission, which extends down to timescales as short as a few seconds,
as evidence for a common
acceleration site with simultaneous injection of electrons both in
low-lying magnetic
features where they produce hard X-rays, and in larger scale and higher
magnetic structures where the radio emission is produced.
\citet{YPW06} discuss another flare in which
Type~III bursts are seen for a long period coincident with HXR emission,
but the locations of the Type III sources are found to be a large
distance from the flare site, at the edge of an associated CME\index{radio emission!type III burst!association with CME}, and the
connection to the HXR emission is less clear.

\begin{figure}[t]
\centerline{\includegraphics[width=0.80\textwidth]{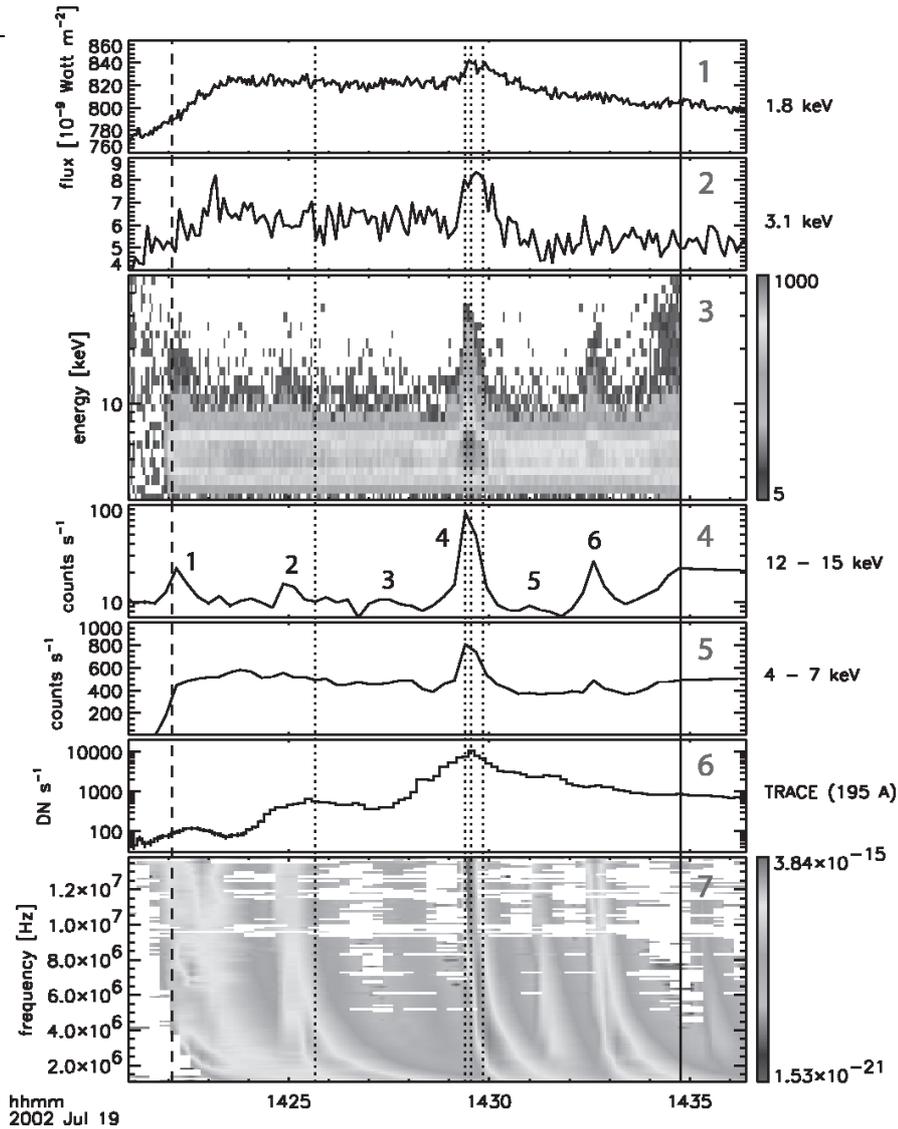}}
\caption{Plot of a group of Type~III bursts (SOL2002-07-19T14:30). 
Panels 1 and 2: \textit{GOES} 1-8 \AA\ (1.6~keV) and 0.5-4 \AA\ (3.1~keV) 
light curves in a linear scale. Panel 3: \textit{RHESSI} spectrogram plot, nighttime
background-subtracted.
Panel 4: \textit{RHESSI} 12-15~keV light curve (nonthermal). Panel 5: \textit{RHESSI} 
4-7~keV light curve (thermal). Panel 6: \textit{TRACE} light  curve integrated over the
active region from which the HXRs originated.
Panel 7: Radio spectrogram from the \textit{WIND}/WAVES instrument. The
dashed line delimits the end of eclipse for \textit{RHESSI}. The solid line
indicates passage through
the South Atlantic Anomaly during which \textit{RHESSI} data is unavailable. From
\citet{CKL08}.}
\index{flare (individual)!SOL2002-07-19T14:30  ($<$B2)!illustration}
\index{radio emission!type III burst!illustration}
\label{fig:white-III}
\end{figure}
\index{satellites!WIND@\textit{WIND}}
\index{satellites!TRACE@\textit{TRACE}}

\citet{CKL08} investigated a group of 6 interplanetary Type III bursts
that showed associated HXR emission at \textit{RHESSI}. Figure~\ref{fig:white-III} shows 
\textit{RHESSI} data during a group of Type III bursts
observed by the \textit{WIND}/WAVES instrument from 2-14~MHz; there is a clear
correlation between the HXR peaks and the onset of Type~III radio
emission for this group\index{hard X-rays!and type III bursts}\index{radio emission!type III burst}.
\citet{CKL08}  discuss (quantitatively) the possibility that the HXRs are
produced by the Type III-emitting electrons via thin-target
bremsstrahlung\index{bremsstrahlung!thin target!type III burst} 
in the corona, and conclude that the number of electrons
required in the Type~III beam would be too large to be consistent with
the observations. Instead, these events look like HXR microflares, with
the HXR spectra being very soft, and the Type III bursts produced during
the energy release in the same way that they are at the onset of
impulsive flares.\index{microflares!and type III bursts}
The energy releases in the events shown in
Figure~\ref{fig:white-III} are less than 10$^{27}$~erg. 

\begin{figure}[t]
\centerline{\includegraphics[width=0.95\textwidth]{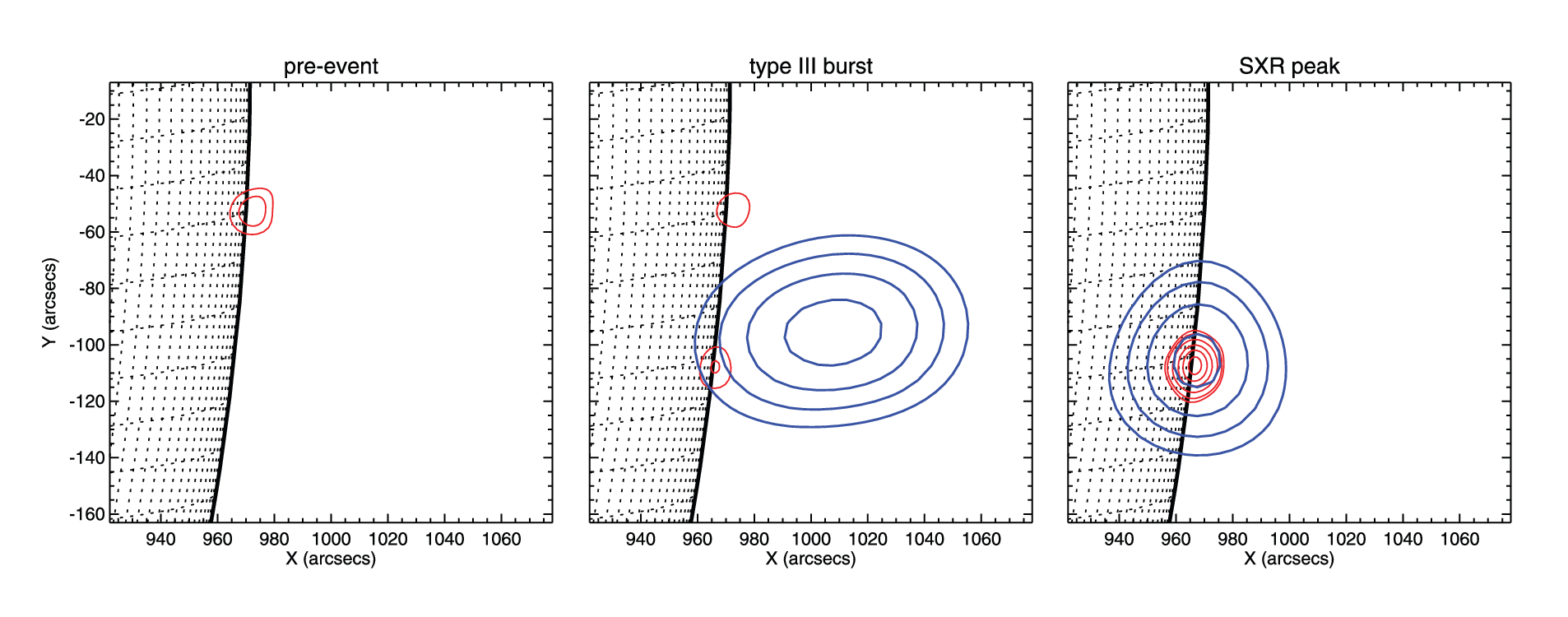}}
\caption{\textit{RHESSI} X-ray imaging before (left), during
(middle), and after (right) radio Type~III bursts. Thermal
emission (4-8~keV) is shown in red contours (contours at 5, 10, 30, 50,
70, 90\% of the level in the final panel) with 10\arcsec\ FWHM resolution,
and the 14-30~keV emission is given in blue (levels are 65, 75, 85, and 
95\%) with a resolution of 60\arcsec. From \citet{KSC08}.
}
\index{radio emission!type III burst!illustration}
\label{fig:white-limbIII}
\end{figure}

On the other hand, \citet{KSC08} discuss a larger event that occurred
at the solar limb such that the main footpoint HXR sources are occulted
and \textit{RHESSI} could easily determine the height of
the coronal HXR sources\index{occulted sources}\index{radio emission!type III burst!and hard X-rays}\index{hard X-rays!and type III bursts}.
They found that the event did show HXR sources at the
solar surface, but when Type III radio emission was present there was a
coronal HXR source\index{coronal sources!radial elongation} elongated in the radial direction (see Figure~\ref{fig:white-limbIII}), consistent with
expectations for thin-target HXRs from the Type III-emitting
electrons. The \textit{WIND} spacecraft was able to measure the number of
nonthermal electrons in the Type III burst reaching the Earth and again it 
appeared to be an order of magnitude too low to explain the observed HXR
flux. These quantitative comparisons are fraught with assumptions that
need to be made but cannot easily be tested: for example, we must 
assume that \textit{WIND}
measures the bulk of the Type III electrons in the solar wind when it
might be merely sampling the edge of the electron beam, and we make assumptions about the
pitch-angle distribution of the nonthermal electrons in the HXR source,
how long the electrons spend in the source, and how open magnetic field lines
connected to the surface diverge with distance in the solar wind. 
These assumptions
introduce uncertainties in the result, but at this point we can say that
there are no cases where we can quantitatively attribute the HXR to the
Type III-emitting electrons. 
On the other hand, based on this event we
can expect that there are more cases of partially-occulted flares
exhibiting radially-elongated HXR emission in conjunction with Type III
bursts waiting to be identified in the \textit{RHESSI} archive.\index{occulted sources}

\begin{figure}[t]
\centerline{\includegraphics[width=0.95\textwidth]{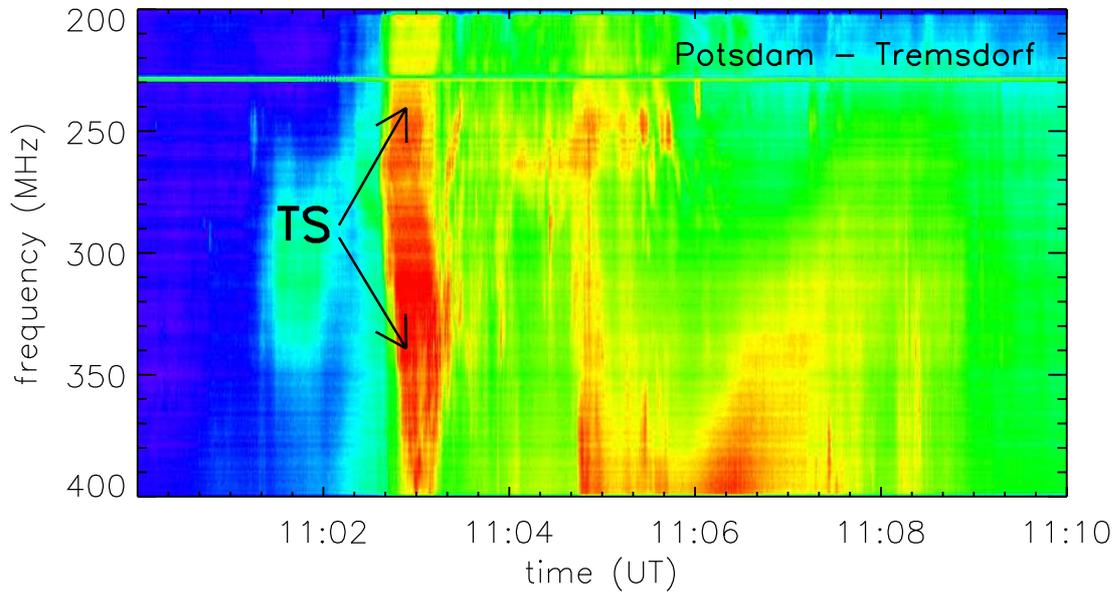}}
\caption{The dynamic spectrum of SOL2003-10-28T11:10 (X17.2), from the
Tremsdorf spectrograph operated by the Astrophysikaliches Institut
Potsdam. 
\index{observatories!Tremsdorf}
The bright chevron-shaped feature centered at 11:03 UT
(labelled ``TS'') is identified as a termination-shock feature. 
}
\label{fig:white-potsdam}
\index{flare (individual)!SOL2003-10-28T11:10 (X17.2)!illustration}
\index{flare (individual)!SOL2003-10-28T11:10 (X17.2)!radio dynamic spectrum}
\index{shocks!termination!illustration}
\end{figure}

\subsection{Acceleration sites in the corona}

As noted above the combination of radio and HXR data is a valuable tool
for understanding the location and nature of energy release sites: the
radio data, in particular, identify the density in the accelerator if
radio emission at the local plasma frequency is visible. This principle
has been applied by \citet{AMR06} in identifying a particular feature in
radio dynamic spectra with the termination shock of a reconnection
outflow.\index{reconnection!outflow}
The argument is that reconnection sites produce
bulk outflows at the Alfv{\'e}n speed and these outflows are likely to
terminate at a shock where the bulk flow energy in the ions can be
converted into heat. Such a shock can also accelerate electrons and thus 
lead to observable plasma emission \citep{ASR02,AuM04}. 
Figure \ref{fig:white-potsdam}
shows a slow-drift feature in the dynamic radio spectrum of SOL2003-10-28T11:10 (X17.2) that \citet{AMR06} interpret as such a termination shock. 
In this event, \NRH\ observations 
above 300~MHz (at frequencies interpreted as harmonic plasma emission) show a
source at the time of the termination-shock feature that is located 0.3
R$_\odot$ (in projection onto the sky) 
away from the flare site \citep[the physical distance may be
larger since the flare is close to disk center; ][]{PMK05}.\index{flare (individual)!SOL2003-10-28T11:10 (X17.2)!termination shock}\index{shocks!termination}\index{termination shock!radio evidence for}
The radio termination-shock feature is seen at the same time that intense $\gamma$-ray
emission is seen, and \citet{AMR06} argue that the termination shock is
the acceleration site for the relativistic $\gamma$-ray-emitting 
electrons. \citet{WMA09} and 
\citet{MWA09} investigate shock-drift acceleration as a means by which 
the energy in such outflows may be converted into electron energy at a
termination shock, utilizing the relativistic theory for shock-drift
acceleration developed by \citet{MAW06}\index{acceleration!shock drift!relativistic}.
As yet there is no
independent confirmation of
the presence either of reconnection outflows or a ``wall'' in the
low-density corona suitable to terminate such a flow, so the true
nature of spectral features such as that in Figure~\ref{fig:white-potsdam}
continues to be a topic for research.  Additional discussion of this 
topic may be found in \citet{HAA10}.

This model is not alone\index{caveats!ambiguity of standard model}\index{standard model!non-uniqueness} in placing the acceleration site for very
energetic electrons that produce observable HXRs and $\gamma$-rays at a
considerable height such that the ambient density 
(in this case, with the fundamental plasma
frequency being around 150~MHz in the termination shock region, the 
electron density is less than $3 \times 10^8$~\pccm) and the magnetic field 
in the acceleration site are both relatively low.\index{frequency!termination shock}
In such models, acceleration takes place in a
relatively low-energy-density region, and then the energetic
electrons must propagate a large distance back to a region of higher energy
density at the flare site low in the atmosphere to produce observable
bremsstrahlung. If the
$\gamma$-ray sources consist of footpoints on either side of a neutral
line, then presumably the oppositely-directed field lines from the
active region must also thread the termination shock despite the large
distance between the acceleration site and the HXR/$\gamma$-ray sources
\citep[on the other hand, non-footpoint locations for the $\gamma$-ray
sources, as for the 2.223~MeV neutron-capture 
line, may point to an acceleration site that 
is not threaded by loops from the active region; ][]{HKL06}. 
These
complications apply to any model in which acceleration does not take
place low in the corona over an active region, and distinguishing
between these two classes of models remains an important step in
identifying the energy release and acceleration mechanisms operating 
in solar flares.

\section{Summary}\label{sec:white-8}

Progress during the \textit{RHESSI} era has confirmed the value of utilizing
radio and HXR data simultaneously to study accelerated electrons in
the solar corona. The complementary nature of the diagnostics in the two
wavelength regimes allows one to study phenomena from differing
viewpoints. However, we have not yet solved several very important
problems, notably the discrepancy between the nonthermal electron 
energy spectral index
derived from radio and HXR observations of electrons which, since they
show the same temporal behavior, must be related; and the anomalous
rising-spectrum submillimeter component. Both of these problems need
high-frequency radio data ($>$30~GHz) for proper study.

\begin{acknowledgements} 
This paper is dedicated to the memory of Mukul Kundu, who passed away
during revision. Mukul was a pioneer of solar radiophysics and one of
the first to realize the close association of radio and hard X-ray
emission and the importance of this relationship for advancing our
understanding of flares.
We thank the organizers of the \textit{RHESSI} Workshop series (Gordon Emslie
and Brian Dennis in particular) and the local organizing committees for each of the
workshops for their hospitality and great efforts to make the workshop 
series such a success.  
Two referees and Bob Lin provided thoughtful comments 
that helped to improve the manuscript. 
Research using \textit{RHESSI} and radio observations was
supported at the University of Maryland by NSF grant ATM 02-33907 and 
NASA contracts NAG 5-12860, NNG-05-GI91G, and NNX-06-AC18G.
\end{acknowledgements} 


\bibliographystyle{ssrv}

\printindex

\end{document}